\documentclass[aps,nofootinbib,prd,notitlepage,superscriptaddress]{revtex4-1}
\usepackage[dvips]{graphicx}
\usepackage{amsmath,amssymb,amsopn,bm}
\usepackage{color}
\usepackage{epstopdf}
\usepackage[normalem]{ulem}
\usepackage{subfig}
\usepackage{verbatim}
\usepackage{bbm}
\usepackage{float}
\usepackage[export]{adjustbox}
\newcommand\myatop[2]{\genfrac{}{}{0pt}{}{#1}{#2}}

\newcommand{\eg}{{\it e.g.}}

\newcommand{\ie}{{\it i.e.}}

\newcommand{\be}{\begin{equation}}
\newcommand{\ee}{\end{equation}}
\newcommand{\bea}{\begin{eqnarray}}
\newcommand{\eea}{\end{eqnarray}}
\newcommand{\bef}{\begin{figure}}
\newcommand{\eef}{\end{figure}}
\newcommand{\bce}{\begin{center}}
\newcommand{\ece}{\end{center}}

\newcommand{\lsim}{\lesssim}
\newcommand{\gsim}{\gtrsim}
\newcommand{\Tpc}{T_{\rm pc}}
\newcommand{\Teff}{T_{\rm eff}}

\begin{document}

\title{Baryonic Sources of Thermal Photons}

\author{Nathan P. M. Holt}
\email{nholt@piedmont.edu}
\affiliation{Piedmont College, Demorest, Georgia, 30535, USA }
\affiliation{Cyclotron Institute and Department of Physics\,\&\,Astronomy,
Texas A{\&}M University, College Station, Texas 77843, USA }
\author{Ralf Rapp}
\email{rapp@comp.tamu.edu}
\affiliation{Cyclotron Institute and Department of Physics\,\&\,Astronomy,
Texas A{\&}M University, College Station, Texas 77843, USA }

\date{\today}

\begin{abstract}
Thermal radiation of photons and dileptons from hadronic matter plays an essential role in understanding electromagnetic
emission spectra in high-energy heavy-ion collisions. In particular, baryons and anti-baryons have been found to be
strong catalysts for electromagnetic radiation, even at collider energies where the baryon chemical potential is small. Here,
we conduct a systematic analysis of $\pi$- and $\omega$-meson-induced reactions off a large set of baryon states. The
interactions are based on effective hadronic Lagrangians where the parameters are quantitatively constrained by empirical
information from vacuum decay branchings and scattering data, and gauge invariance is maintained by suitable
regularization procedures. The thermal
emission rates are computed using kinetic theory but can be directly compared to previous calculations using hadronic
many-body theory. The comparison to existing calculations in the literature reveals our newly identified contributions to be
rather significant.
\end{abstract}

\maketitle

\section{Introduction}
\label{sec:intro}
Electromagnetic (EM) radiation from the fireballs formed in heavy-ion collisions offers a wide range of insights into the
properties of matter governed by Quantum Chromodynamics (QCD). Low-mass dilepton spectra, at invariant masses
$M\lsim1$\,GeV, directly probe
how the spectral functions (SFs) of vector mesons (most notably the $\rho$-meson) transit from massive and confined degrees
of freedom in the QCD vacuum into a rather structureless spectrum at high temperature and density, suggestive of a
quark-antiquark continuum~\cite{Rapp:1999ej}. At intermediate masses, $M\gsim1.5$\,GeV, continuum radiation is
chiefly emitted from the early phases of the fireball, and its inverse slope serves as an excellent thermometer of the
medium~\cite{Arnaldi:2008er}.  While invariant-mass spectra are, by definition, unaffected by any Doppler blueshift
caused by the collective expansion of the fireball, this is no longer the case for
the transverse-momentum ($q_T$) spectra of both photons and dileptons. Temperature extractions from the $q_T$ spectra
therefore require a deconvolution of the radial flow of the emitting fireball cells. The spectra of ``direct" photons (obtained
after subtracting long-lived final-state hadron decays) measured at RHIC and LHC not only exhibit excess yields indicating a
robust signal of thermal radiation, but also carry a significant asymmetry in the azimuthal emission angle ($\phi$) in the
transverse plane, commonly associated with the ``elliptic flow" of the underlying hydrodynamic medium. Measurements of the
pertinent coefficient, $v_2(q_T)$,  of the $\cos(2\phi)$ modulation in the photon $q_T$ spectra reach values which are
not far from those observed for pions. The latter,
however, are only emitted at the end of the fireball evolution, \ie, at the kinetic decoupling of the hadrons which typically
occurs at a freezeout temperature of around $T_{\rm fo}\simeq 100$~MeV. This suggests that many of the photons are
radiated well into the fireball evolution, where most of the medium's $v_2$ has already built up. The typical timescale for hydrodynamic simulations to achieve this in mid-central collisions of heavy nuclei is about 5~fm/$c$, at which point the
medium has cooled down to temperatures near the pseudocritical one, $\Tpc\simeq 160$\,MeV. On the other hand,
the experimentally measured inverse-slope parameters amount to about
$\Teff\simeq (240\pm30)$\,MeV  in 0.2\,TeV Au-Au collisions at RHIC~\cite{Adare:2014fwh} and
$\Teff\simeq (300\pm40)$\,MeV  in 2.76\,TeV Pb-Pb collisions at the LHC~\cite{Adam:2015lda}. The former is consistent
with thermal emission over a broad window around $\Tpc$ once the blueshift effect is included~\cite{vanHees:2011vb},
$\Teff \simeq T \sqrt{(1+\beta_{\rm avg})/(1-\beta_{\rm avg})}$, with an average radial flow velocity of
$\beta_{\rm avg}\simeq1/3$, yielding $\Teff\simeq230$\,MeV,  while the LHC results possibly indicate somewhat
 higher local emission temperatures. These results suggest the photon emissivity in hot hadronic matter as a key ingredient
to interpret the data~\cite{Turbide:2003si,vanHees:2011vb}. One may argue that the present
understanding of the direct-photon data at RHIC and the LHC is not yet complete, as state-of-the-art
calculations~\cite{vanHees:2014ida,Paquet:2015lta} still fall somewhat short of the experimental results, both in
spectral yields and $v_2$, at the  1-2~$\sigma$ level. It is thus of interest to further scrutinize the thermal photon
emissivities of QCD matter.

The photon polarization tensor needed to compute the thermal emission rate is continuously connected to that of low-mass
dileptons via  the $M\to0$ limit of the latter at finite three-momentum, $q$. The low-mass dilepton excess is known to
receive important contributions from baryonic sources~\cite{Rapp:1999ej} even at the small baryon chemical potentials
created in the mid-rapidity region at collider energies~\cite{Rapp:2000pe}. This is due to the sum of rate contributions from baryons and anti-baryons, together with their total number being considerable at hadrochemical freezeout at RHIC and the LHC.
Of particular interest for the production rate of photons at
phenomenologically relevant energies, $q_0\simeq 1$\,GeV (which, as mentioned above, get blueshifted to higher $q_T$ in
the measured spectra), are  $t$-channel exchange reactions (\eg, $\pi$ exchange in $\pi N\to\gamma N$), since they do
not suffer a $1/q_0^2$ suppression as do the resonant production channels (\eg, $\pi N \to \Delta \to \gamma N$). In the
language of hadronic many-body theory, the $t$-channel production processes correspond to medium modifications of the
virtual meson cloud of the photon, or, within the vector meson dominance model (VDM), to the meson cloud of
the light vector mesons (mostly the $\rho$). In cold nuclear matter, the pion cloud modifications of the $\rho$ meson have
been well constrained~\cite{Urban:1998eg,Rapp:1999ej}, but the effects due to thermally excited baryons have thus far
been treated in an approximate way, by introducing an effective nucleon density~\cite{Rapp:1999us}. In the present
manuscript we elaborate on this approximation with an explicit calculation using a rather extensive set of baryon resonance
states, $B_i$, for $\pi B_1B_2$ couplings where the corresponding vertices are constrained by scattering data and
empirical decay branchings. Furthermore, guided by the importance of the $\pi\rho\omega$ coupling as found in our
previous work on photon rates from a meson gas~\cite{Holt:2015cda}, we extend these calculations to the baryonic
sector by including the $\pi\omega$ cloud of the $\rho$ meson corresponding to $\omega$ ($\pi$) $t$-channel exchange
reactions in $\pi B_1\to\rho B_2$ ($\omega B_1\to\gamma B_2$) reactions. We compute the pertinent production
rates with the standard kinetic-theory expression and conduct quantitative comparisons to existing rates from the
in-medium $\rho$ SF.

This article is organized as follows. In Sec.~\ref{sec:micro} we lay out the microscopic ingredients of our model, by first
introducing the hadronic interaction Lagrangians (Sec.~\ref{ssec:lagrangian}) followed by a discussion of the
phenomenological vertex formfactors (Sec.~\ref{ssec:ff}) and the evaluation of the adjustable parameters
(Sec.~\ref{ssec:para}). In Sec.~\ref{sec:picloud} we report the results for the energy dependent photon rates
classified into contributions from pion-baryon $S$-wave (Sec.~\ref{ssec:swave}), $P$-wave (Sec.~\ref{ssec:pwave}
and $D$-wave (Sec.~\ref{ssec:dwave}) interactions and their total in comparison to previous calculations.
In Sec.~\ref{sec:pi-om-cloud} we discuss baryon-induced photon rates involving the $\pi\rho\omega$ vertex, classified
into processes with internal $\omega$ exchanges (Sec.~\ref{ssec:omexch}) and with in- or outgoing $\omega$ mesons
(Sec.~\ref{ssec:omext}). In Sec.~\ref{sec:results} we give an overall assessment of how our newly calculated rates
figure in the context of existing calculations, both in the net-baryon free region relevant for collider energies and for
moderate baryon chemical potentials. We summarize, conclude, and give an outlook in Sec.~\ref{sec:concl}.

\section{Hadronic Lagrangians and Parameter Constraints}
\label{sec:micro}
Our calculations of the thermal photon emission rate from hot hadronic matter will be based on the standard kinetic-theory
expression for production channels of the type $h_1 + h_2 \to h_3 + \gamma$ ($h_i$: hadrons),
\begin{equation}
\begin{split}
\label{eq:rate}
q_0 \frac{dR_{\gamma}}{d^3 q} = & \mathcal{N} \int \frac{d^3 p_1}{(2 \pi)^3 2E_1} \frac{d^3 p_2}{(2 \pi)^3 2E_2}
\frac{d^3 p_3}{(2 \pi)^3 2E_3} \, \overline{|{\cal M}_{12\to3\gamma}|^2} \\
& \times (2 \pi)^4 \delta^4 (p_1+p_2-p_3-q) f(E_1,T) f(E_2,T) \frac{[1 \pm f(E_3,T)]}{2(2 \pi)^3} \  ,
\end{split}
\end{equation}
where $\mathcal{N} = \mathcal{N}_1 \mathcal{N}_2$ is the overall degeneracy factor of incoming particles, the $f$'s are Fermi or Bose distribution functions and the ``$\pm$'' is ``+($-$)'' if $h_3$ is a meson (baryon). The key ingredient is the invariant matrix element ${\cal M}_{12\to3\gamma}$ for the photon producing scattering process, which we calculate from suitably defined and constrained hadronic Lagrangians. Throughout this paper, we will invoke the VDM, \ie, all our photon producing reactions will
be channeled through an intermediate $\rho$ meson converting into a photon.

\begin{figure}[H]
\begin{center}
\subfloat[$s$-channel]{
\includegraphics[scale=0.75]{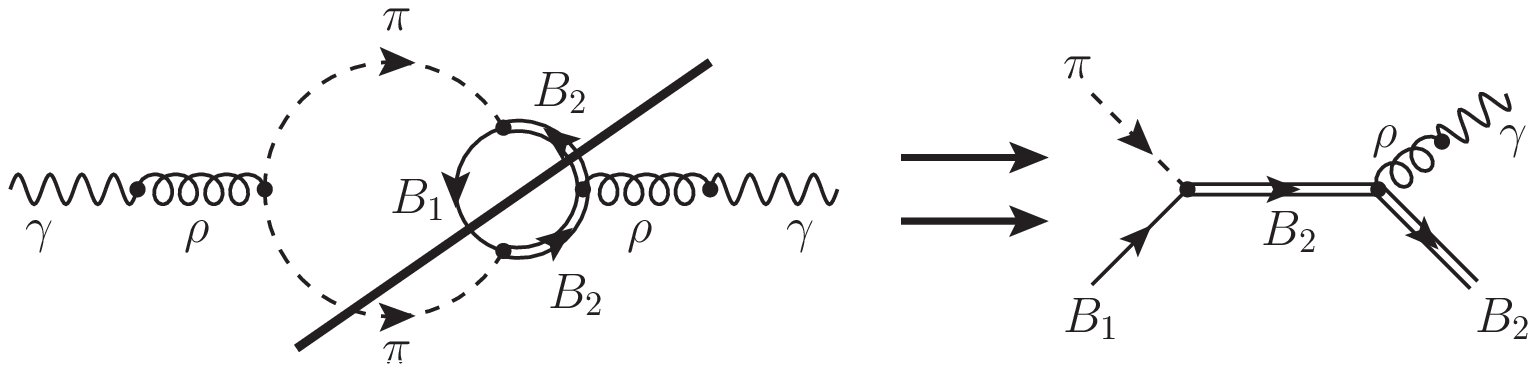}}

\subfloat[$t$-channel]{
\includegraphics[scale=0.75]{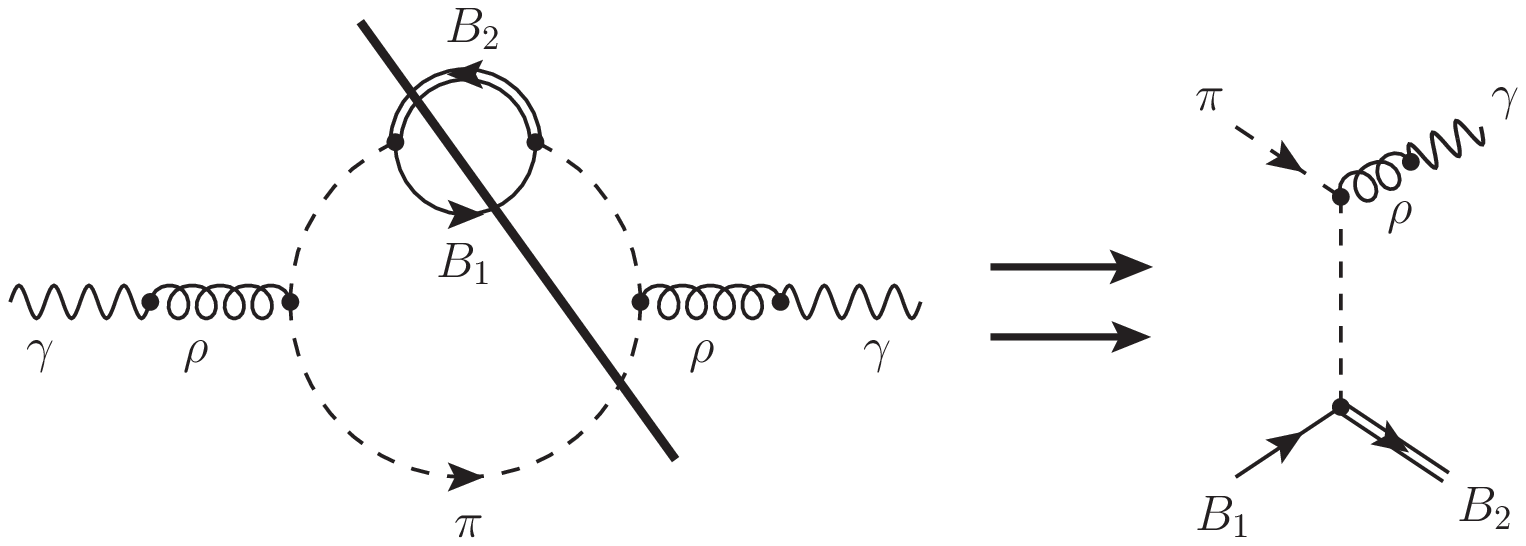}}

\subfloat[$u$-channel]{
\includegraphics[scale=0.75]{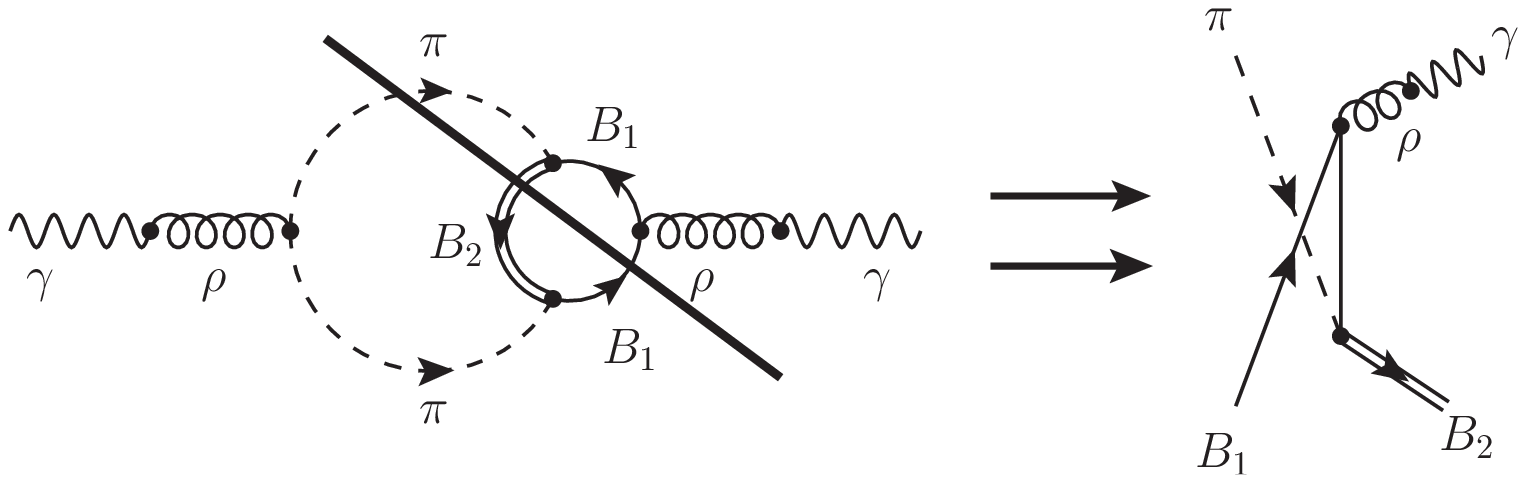}}

\subfloat[contact term]{
\includegraphics[scale=0.75]{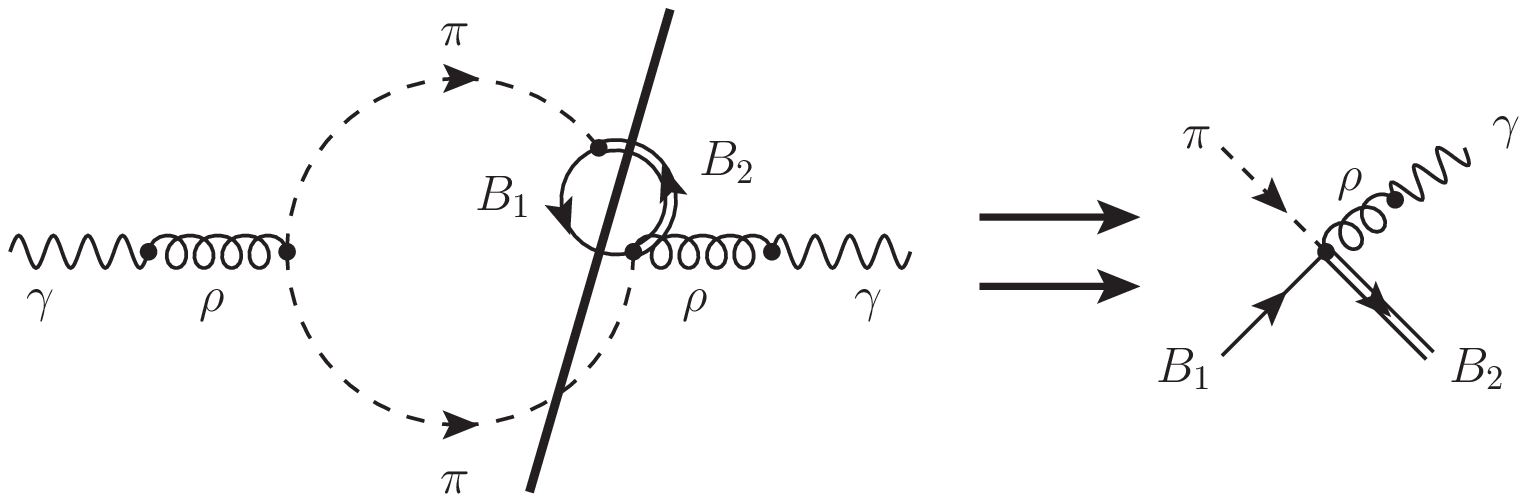}}
\end{center}
\caption{Cuts to pion cloud modifications of the in-medium $\rho$ selfenergy which yield Born scattering diagrams considered in this work. Vertex corrections and internal-propagator dressing give rise to $s$-, $t$- and $u$-channel diagrams as well as 4-point interactions dictated by gauge invariance.}
\label{fig:pi-cloud-cuts}
\end{figure}

\begin{figure}[H]
\begin{center}
\subfloat[``Internal'' $\omega$]{
\includegraphics[scale=0.75]{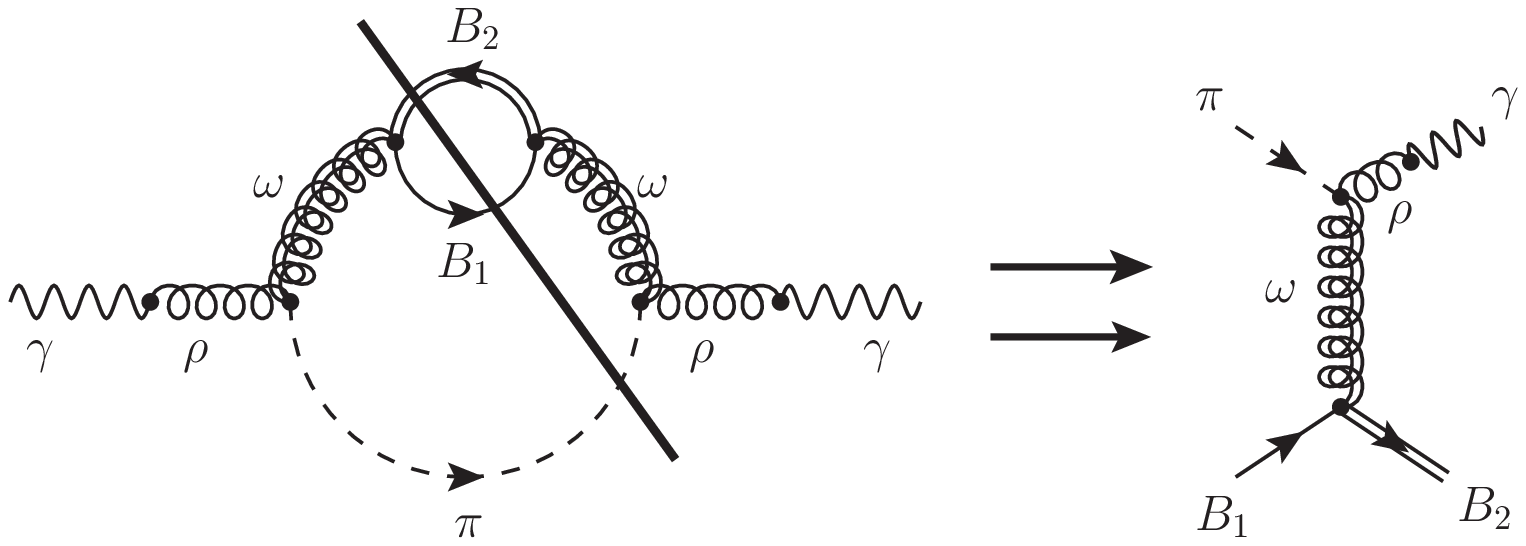}}

\subfloat[``External'' $\omega$]{
\includegraphics[scale=0.75]{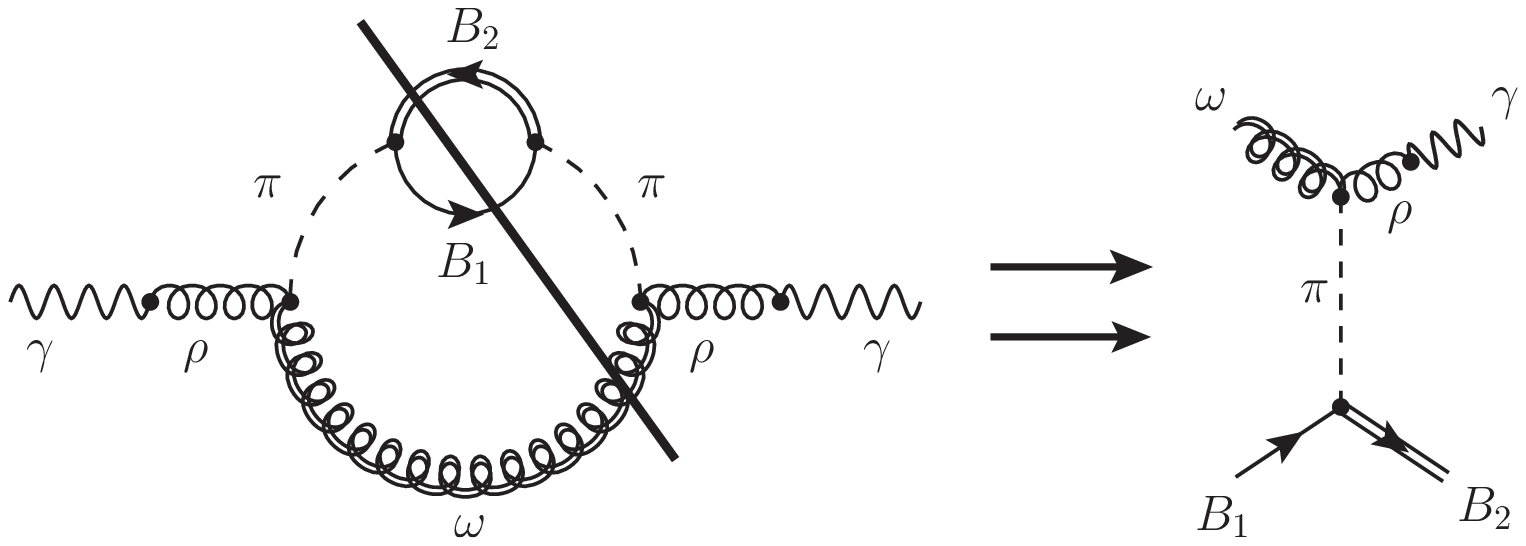}}
\end{center}
\caption{Cuts to the $\pi\omega$ cloud of the $\rho$ selfenergy giving rise to processes involving an ``internal'' $\omega$ in the scattering process (top row), and to an ``external'' $\omega$ (bottom row).}
\label{fig:omega-cloud-cuts}
\end{figure}

Let us briefly recall the relation of the kinetic-theory framework to many-body
calculations of the in-medium $\rho$ selfenergy, to which we will refer on several occasions throughout this paper. Within the
VDM, the cuts (imaginary parts) of the two-loop $\rho$ selfenergy directly correspond to Born-level scattering diagrams for photon production figuring in the kinetic-theory expression, Eq.~(\ref{eq:rate}), as illustrated in Figs.~\ref{fig:pi-cloud-cuts} and
\ref{fig:omega-cloud-cuts}. The outer loop,  $\pi\pi$ or $\pi\omega$, generates the imaginary part of the (timelike) $\rho$ selfenergy in vacuum, while a further in-medium loop on the pion or omega propagators (or medium-induced vertex corrections) produces a non-vanishing imaginary part at the photon point,
$M \to 0$. This connection is also the origin of referring to pion cloud or
$\pi\omega$ cloud contributions to the selfenergy.

The remainder of this section is organized as follows. In Sec.~\ref{ssec:lagrangian} we introduce the effective Lagrangians used in this work, where we first lay out the fully relativistic versions (Sec.~\ref{sssec_rel}) followed by a non-relativistic reduction for baryons
(Sec.~\ref{sssec_nonrel}). In Sec.~\ref{ssec:ff} we implement hadronic vertex formfactors which simulate finite-size effects and are essential for quantitative applications to phenomenology; specifically, we discuss the pion-induced reactions with baryons (Sec.~\ref{sssec_piB}), mesonic interactions involving the $\pi\rho\omega$ vertex (Sec.~\ref{sssec_piroom}), and
$\omega$-induced interactions with baryons (Sec.~\ref{sssec_omB}), while taking special care to maintain electromagnetic gauge invariance.
In Sec.~\ref{ssec:para}, our procedures for fixing the parameters are elaborated upon, starting with coupling constants estimated
from baryon resonance decay branching ratios (Sec.~\ref{sssec_cc}) and followed by evaluating constraints for the formfactor cutoffs by using nuclear photoabsorption cross sections  (Sec.~\ref{sssec_photoabs}) and $\pi N$ scattering data
(Sec.~\ref{sssec_piN}).

\subsection{Effective Lagrangians}
\label{ssec:lagrangian}

\subsubsection{Relativistic Interaction Lagrangians}
\label{sssec_rel}
We use the notation $a B_1 B_2$ to indicate an interaction between a meson $a$ and two baryons $B_1$ and $B_2$, which may or may not be the same particle species.
We begin with the free-field Lagrangian terms for $\pi$ mesons, $\rho$ mesons, and massive spin-$1/2$ baryons:
\begin{align}
\label{eq:free}
\mathcal{L}_{B_{\frac{1}{2}}} &= \bar{\psi} \left( i \gamma^{\mu}\partial_{\mu} -m_B \right) \psi \, , \\
\mathcal{L}_{\pi} &= \frac{1}{2}\partial_{\mu} \vec{\pi} \cdot \partial^{\mu}
\vec{\pi} - \frac{1}{2} m_{\pi}^{2} \vec{\pi} \cdot \vec{\pi} \, , \\
\mathcal{L}_{\rho} &= -\frac{1}{4} \vec{\rho}_{\mu \nu} \cdot \vec{\rho}^{\,\mu \nu}
+ \frac{1}{2} m_{\rho}^{2} \vec{\rho}_{\mu} \cdot \vec{\rho}^{\, \mu} \, ,
\end{align}
 where the notation $B_{\frac{1}{2}}$ indicates a spin-1/2 baryon, and the $\rho$ field strength tensor is
\begin{equation}
\vec{\rho}_{\mu \nu} = \partial_{\mu} \vec{\rho}_{\nu} -
\partial_{\nu} \vec{\rho}_{\mu} \ .
\end{equation}
 To describe spin-3/2 baryons we use the Rarita-Schwinger formalism~\cite{Rarita:1941mf}, where the free-field Lagrangian for a massive spin-3/2 particle is given by
\begin{equation}
\label{eq:freeRS}
\mathcal{L}_{B_{\frac{3}{2}}}= - \bar{\psi}_{\mu} \left( i \gamma^{\mu}\partial_{\mu} -m_B \right) \psi^{\mu}
+ \frac{i}{3} \bar{\psi}_{\mu} \left( \gamma^{\mu} \partial_{\nu} + \gamma_{\nu}\partial^{\mu} \right) \psi^{\nu}
- \frac{1}{3} \bar{\psi}_{\mu} \gamma^{\mu} \left( i \gamma^{\mu}\partial_{\mu} +m_B \right) \gamma_{\nu} \psi^{\nu} \, .
\end{equation}
For a $\pi B_1 B_2$ interaction term with two spin-1/2 baryons, we choose a derivative coupling to respect chiral symmetry~\cite{Weinberg:1996kr},
\begin{equation}
\label{eq:piBB-N}
\mathcal{L}_{\pi B_{ \frac{1}{2} \frac{1}{2} }} = \frac{f_{\pi B_1 B_2}}{m_{\pi}} \bar{\psi} ( \gamma_5) \gamma^{\mu}\partial_{\mu} \vec{\pi} \cdot \vec{\mathcal{T}} \psi \, .
\end{equation}
Since the pion field is parity-odd, a $\gamma_5$ factor is needed (or not) if both baryon fields are of the same (different) internal parity. In the above expression $\vec{\mathcal{T}}$ is the pion isospin transition operator acting on the baryon fields.  For couplings between pions and two spin-3/2 baryons we make the ansatz
\begin{equation}
\label{eq:piBB-D}
\mathcal{L}_{\pi B_{ \frac{3}{2} \frac{3}{2} }} =  \frac{f_{\pi B_1 B_2}}{m_{\pi}} \bar{\psi}_{\mu} \left( \gamma_5 \right) \gamma^{\nu} \partial_{\nu} \vec{\pi} \cdot \vec{\mathcal{T}} \psi^{\mu} \, ,
\end{equation}
with the same requirement for inclusion of the $ \gamma^5$ term as in Eq.~(\ref{eq:piBB-N}). Interactions between pions, spin-1/2, and spin-3/2 particles are given by~\cite{Urban:1998eg}
\begin{equation}
\mathcal{L}_{\pi B_{ \frac{1}{2} \frac{3}{2} }} = -\frac{f_{\pi B_1 B_2}}{m_{\pi}} \bar{\psi}_{\mu} \left( \gamma_5 \right)
\partial^{\mu} \vec{\pi} \cdot \vec{\mathcal{T}} \psi + \mathrm{H.c.} \, ,
\end{equation}
where ``H.c.'' indicates the Hermitian conjugate of the previous term.

To obtain a non-relativistic $D$-wave interaction we use the following interaction term between pions, spin-1/2, and spin-3/2 baryons with differing parity quantum numbers~\cite{Gasparyan:2003fp}:
\begin{equation}
\label{eq:Dwaverel}
\mathcal{L}^D_{\pi B_{ \frac{1}{2} \frac{3}{2} }} = \frac{f_{\pi B_1 B_2}}{m_{\pi}^2} \bar{\psi}^{\mu} \gamma_5 \gamma^{\nu} \partial_{\nu} \partial_{\mu}
\vec{\pi} \cdot \vec{\mathcal{T}} \psi + \mathrm{H.c.} \, .
\end{equation}
The parity-violating $\pi \rho \omega$ interaction is incorporated using the Wess-Zumino term~\cite{Wess:1971yu,Witten:1983tx},
\begin{equation}
\label{eq:WZ}
\mathcal{L}_{\pi\rho\omega} = g_{\pi\rho\omega} \epsilon^{\mu \nu \alpha \beta}
\partial_{\alpha} \omega_{\beta} \partial_{\mu} \vec{\rho}_{\nu}
\cdot \vec{\pi} \ .
\end{equation}
We generate interactions with the $\rho$ meson by applying the process of minimal substitution to the above Lagrangians,
\begin{equation}
\partial_{\mu} \to \partial_{\mu} + i g_{\rho} \vec{\rho}_{\mu} \cdot \vec{\mathcal{T}} \, ,
\end{equation}
where $g_{\rho}$ is the isospin gauge coupling of the $\rho$. Owing to vector meson universality~\cite{Sakurai,Ericson:1988gk}, the $\rho$ couples with approximately universal strength to all particles carrying isospin. We identify $g_{\rho}$ with the
$\rho\pi\pi$ coupling, $g_{\rho} = g_{\rho \pi \pi}$. Applying this gauging procedure to the above Lagrangians generates the following interactions:
\begin{align}
\label{eq:RelLagrangians}
\mathcal{L}_{\rho \pi \pi} &= -g_{\rho} \vec{\rho}^{\,\mu} \cdot \left( \partial_{\mu} \vec{\pi} \times \vec{\pi} \right) \, , \nonumber \\
\mathcal{L}_{\rho\rho\rho} &= -\frac{1}{2} g_{\rho} \vec{\rho}^{\,\mu \nu} \cdot
\left(\vec{\rho}_{\mu} \times \vec{\rho}_{\nu} \right) \, ,
\nonumber \\
\mathcal{L}_{\pi \rho \rho \omega} &= g_{\pi\rho\omega} g_{\rho}
\epsilon^{\mu \nu \alpha \beta} \partial_{\alpha} \omega_{\beta}
\left( \vec{\rho}_{\mu} \times \vec{\rho}_{\nu} \right) \cdot \vec{\pi} \, ,
\nonumber \\
\mathcal{L}_{\rho B_{ \frac{1}{2} \frac{1}{2} }} &= -g_{\rho} \bar{\psi} \gamma^{\mu} \vec{\rho}_{\mu} \cdot \vec{\mathcal{T}} \psi \, ,
\nonumber \\
\mathcal{L}_{\rho B_{ \frac{3}{2} \frac{3}{2} }} &= g_{\rho} \bar{\psi}_{\mu} \gamma^{\nu} \vec{\rho}_{\nu} \cdot \vec{\mathcal{T}}
- \frac{g_{\rho}}{3} \bar{\psi} \gamma^{\mu} \left( \gamma^{\mu}\vec{\rho}_{\nu} + \gamma_{\nu}\vec{\rho}^{\,\mu} \right) \cdot \vec{\mathcal{T}} \psi^{\nu} +
\frac{g_{\rho}}{3} \bar{\psi}_{\mu} \gamma^{\mu}\gamma^{\nu}\vec{\rho}_{\nu} \cdot \vec{\mathcal{T}} \gamma_{\sigma} \psi^{\sigma} \, ,
\nonumber \\
\mathcal{L}_{\pi \rho B_{ \frac{1}{2} \frac{1}{2} } } &= g_{\rho} \frac{f_{\pi B_1 B_2}}{m_{\pi}} \bar{\psi} ( \gamma_5) \left( \gamma^{\mu}
\vec{\rho}_{\mu} \times \vec{\pi} \right) \cdot \vec{\mathcal{T}} \psi \, ,
\nonumber \\
\mathcal{L}_{\pi \rho B_{ \frac{3}{2} \frac{3}{2} }}  &= g_{\rho} \frac{f_{\pi B_1 B_2}}{m_{\pi}} \bar{\psi}_{\mu} ( \gamma_5) \left( \gamma^{\nu}
\vec{\rho}_{\nu} \times \vec{\pi} \right) \cdot \vec{\mathcal{T}} \psi^{\mu} \, ,
\nonumber \\
\mathcal{L}_{\pi \rho B_{ \frac{1}{2} \frac{3}{2} } } &= -g_{\rho} \frac{f_{\pi B_1 B_2}}{m_{\pi}} \bar{\psi}^{\mu} ( \gamma_5) \left( \vec{\rho}_{\mu} \times \vec{\pi} \right) \cdot \vec{\mathcal{T}} \psi + \mathrm{H.c.} \, ,
\nonumber \\
\mathcal{L}^D_{\pi \rho B_{ \frac{1}{2} \frac{3}{2} } } &= g_{\rho} \frac{f_{\pi B_1 B_2}}{m_{\pi}^2} \bar{\psi}^{\mu}
\left( \gamma_5 \right) \gamma^{\nu}
\left( \vec{\rho}_{\nu} \times \partial_{\mu}\vec{\pi} + \vec{\rho}_{\mu} \times \partial_{\nu} \vec{\pi} \right)
\cdot \vec{\mathcal{T}} \psi + \mathrm{H.c.} \, .
\end{align}
Since the $\omega$ is an isosinglet vector meson, we model the $\omega B_1 B_2$ interaction to be similar in structure to the $\rho B_1 B_2$ vertex~\cite{Machleidt:1989tm,Bieniek:2001xu,Gasparyan:2003fp}:
\begin{align}
\label{eq:omegaNN}
\mathcal{L}_{\omega B_{ \frac{1}{2} \frac{1}{2} } } &= g_{\omega B_1 B_2} \bar{\psi} ( \gamma_5) \gamma^{\mu} \omega_{\mu}  \psi \, , \nonumber \\
\mathcal{L}_{\omega B_{ \frac{1}{2} \frac{3}{2} } } &=g_{\omega B_1 B_2} \bar{\psi}^{\mu} ( \gamma_5) \omega_{\mu} \psi + \mathrm{H.c.} \, .
\end{align}
There is no $\vec{\mathcal{T}}$ operator in Eq.~(\ref{eq:omegaNN}) since the $\omega$ is an isospin-0 particle and cannot induce transitions between isospin-1/2 and -3/2 states.
Electromagnetic interactions are modelled using the VDM, where the EM current couples to hadrons exclusively through
neutral vector mesons, \ie, the $\rho$, $\omega$, and $\phi$. In the present work we only account
for the $\rho$ coupling to the photon, as the $\omega$ and $\phi$ couplings are smaller by factors of $\approx$ 11 and
$\approx$ 7, respectively, relative to the $\rho \gamma$ coupling~\cite{Sakurai,Turbide:2003si,Holt:2015cda}. The
latter is given by
\begin{equation}
\label{eq:VMD}
\mathcal{L}_{\rho \gamma} = -A^{\mu} C_{\rho} m_{\rho}^2 \rho^{0}_{\mu} \ ,
\end{equation}
where $\rho^{0}_{\mu}$ is the charge neutral ($I_3=0$) component of the $\rho$ field.
Using strict VDM would give the $C_{\rho}= e/g_{\rho}$. However, as in the previous work~\cite{Holt:2015cda}, we treat
it as an adjustable parameter which can be fitted via the $\rho$ di-electron decay. In practice, these two
values differ by $\approx 15\%$.

In the present work all processes involving the $\omega$, whether as an external or internal particle, will be coupled via
the $\pi \rho \omega$ vertex of Eq.~(\ref{eq:WZ}) with the $\rho$ coupling to an on-shell photon via VDM. The Levi-Civita
tensor structure of the $\pi \rho \omega$ vertex ensures the corresponding Born diagram to be gauge invariant by itself.

Equations~(\ref{eq:piBB-N})-(\ref{eq:VMD}) are the relativistic Lagrangians from which we will derive our non-relativistic interactions.

\subsubsection{Non-Relativistic Reductions}
\label{sssec_nonrel}
To simplify our calculations we exploit the energy scale of baryon masses by expanding Dirac and Rarita-Schwinger spinors to $0^{\mathrm{th}}$ order in momentum, \textit{i.e.} $(|\vec{p}_B|/m_B)^0$. This treatment avoids possible ambiguities in use of the Rarita-Schwinger propagator ~\cite{Velo:1970ur,Benmerrouche:1989uc} and agrees with the non-relativistic interactions used in the works we are augmenting~\cite{Urban:1998eg,Urban:1999im,Rapp:1999us}. Additionally, the non-relativistic treatment of
spin-3/2 particles significantly simplifies calculations while maintaining accuracy at values of three-momenta up to at least
$q \approx 2$ GeV~\cite{Riek:2008ct}.  We also simplify the baryonic propagators by neglecting antiparticle contributions,
but keeping relativistic kinematics in the denominator~\cite{Urban:1998eg}:
\begin{equation}
G_B(p) = \frac{ \underset{\mathrm{spin}}{\sum}\bar{\psi} \psi  }{ p^2 -m_B^2+i\epsilon}
 \to   \frac{1}{p_0 - \omega_B(\vec{p}\,) +i\epsilon} \, ,
\end{equation}
where $\omega_B(\vec{p}\,) = \sqrt{ \vec{p}^{\,2} +m_B^2 }$ is the on-shell energy of the baryon and $p_0$ is its off-shell energy.

The inclusion of the $\gamma_5$ matrix to compensate for baryon parity differences results in differing non-relativistic
interactions. We use the notation $\mathcal{L}^{+}$ to indicate the two baryon spinors to have the same parity quantum
number (+1 or -1) and $\mathcal{L}^{-}$ to indicate they have opposite parity quantum numbers. For the interaction of a
pion with two spin-1/2 or two spin-3/2 baryons the non-relativistic reduction of the Dirac spinors in  Eqs.~(\ref{eq:piBB-N})
and (\ref{eq:piBB-D}) leads to
\begin{align}
\mathcal{L}_{\pi B_1 B_2}^{+} &= \frac{f_{\pi B_1 B_2}}{m_{\pi}} \chi^{\dag}_1 \left(\vec{\pi} \cdot \vec{\mathcal{T}} \right) \left(\vec{k} \cdot \vec{\mathcal{S}}\right) \chi_2 \, , \\
\mathcal{L}_{\pi B_1 B_2}^{-} &=\frac{f_{\pi B_1 B_2}}{m_{\pi}} \chi^{\dag}_1 \left(\vec{\pi} \cdot \vec{\mathcal{T}}
\right) \omega_{\pi}(k) \,  \chi_2 \, .
\label{eq:pi-BBvertex}
\end{align}
The three-momentum of the pion is denoted by $\vec{k}$, and its on-shell energy is
$\omega_{\pi}(k)=\sqrt{m_{\pi}^2+\vec{k}^2}$. Here the $\chi$ are either two- or four-component spinors in both spin and
isospin space, depending on the quantum numbers of the baryons. The $\vec{\mathcal{S}}$ denotes the spin transition operator
connecting the baryon spinors. We see that the $\mathcal{L}^{-}$ yields an $S$-wave interaction, while the $\mathcal{L}^{+}$
gives a $P$-wave interaction. In principle, we could expand the relativistic interactions to higher orders in $(|\vec{p}_B|/m_B)$
to obtain a $D$-wave interaction. However, this would be at variance with a strict leading-order expansion. Therefore, we
perform our non-relativistic reduction on the relativistic Lagrangian of Eq.~(\ref{eq:Dwaverel}).  The non-relativistic spin
structure of the resulting $D$-wave interaction is slightly more complicated than the $P$-wave interaction and depends on
the spin of $\chi_2$, either 1/2 or 3/2. Under the assumption that $\chi_1$ is spin-1/2, the resulting interactions are
\begin{equation}
\label{eq:Dwave}
\mathcal{L}_{\pi B_1 B_2}^{D} =\frac{f_{\pi B_1 B_2}}{m_{\pi}^2} \begin{cases}
\chi^{\dag}_1 \left(\vec{\pi} \cdot \vec{\mathcal{T}} \right)
\left(\vec{k} \cdot \vec{\mathcal{S}} \right) \left(\vec{k} \cdot \vec{\sigma} \right)   \chi_2 \, ,
 \quad \ \chi_2 \hspace{3pt} \mathrm{spin} = \frac{1}{2}
\\
 \chi^{\dag}_1 \left(\vec{\pi} \cdot \vec{\mathcal{T}} \right)
\left(\vec{k} \cdot \vec{\sigma} \right) \left(\vec{k} \cdot \vec{\mathcal{S}}^{\dag} \right)  \chi_2 \, ,
 \quad  \chi_2 \hspace{3pt} \mathrm{spin} = \frac{3}{2} \ .
\end{cases}
\end{equation}
The non-relativistic reduction of baryon interactions with the $\rho$ is similarly straightforward. However, the
$\rho B_1 B_2$ interaction requires modification to satisfy a non-relativistic Ward-Takahashi
identity~\cite{Herrmann:1993za,Urban:1998eg}. The resulting modified interaction vertex is
\begin{equation}
\label{eq:rhoBB}
\mathcal{L}_{\rho B_1 B_2}^{+} = - g_{\rho} \chi^{\dag}_1 \rho_{\mu} \left(
\myatop{ \frac{G_{B_1}^{-1} (p+q) - G_{B_2}^{-1} (p)}{q_0} } {0} \right) ^{\mu}
\left(\vec{\rho} \cdot \vec{\mathcal{T}} \right) \chi_2 \ ,
\end{equation}
where $p$ and $q$ are the baryon and $\rho$ four-momenta, respectively. We use the notation
\begin{equation}
\left( \myatop{\mathbbm{1}}{\vec{\mathcal{S}}} \right)^{\mu} = \begin{cases}
\mathbbm{1} \, , & \mu = 0 \\
\mathcal{S}_{\mu} \, , & \mu = 1,2,3 \,  \end{cases}
\end{equation}
to indicate the four-vector nature of the object in parentheses.
We note that in our framework there exists no interaction between the $\rho$ meson and two different species of baryons.
This is a result of effectively introducing the $\rho$ as a gauge boson via minimal substitution, so that the $\rho B_1 B_2$
interactions generated via gauging the free-field Lagrangians of Eqs.~(\ref{eq:free}) and (\ref{eq:freeRS}) result only in
interactions between baryons of the same type. This implementation of $\rho$ interactions also precludes the possibility of
double-counting direct $\rho B_1 B_2$ interactions which have already been included in the in-medium $\rho$ spectral
function of Refs.~\cite{Urban:1998eg,Urban:1999im,Rapp:1999us}. The non-relativistic baryon contact terms are
\begin{align}
\mathcal{L}^{+}_{\pi \rho B_1 B_2} &= g_{\rho} \frac{f_{\pi B_1 B_2}}{m_{\pi}} \chi^{\dag}_1
\rho_{\mu} \left( \myatop{0}{\vec{\mathcal{S}}} \right)^{\mu}
\left( \vec{\rho} \times \vec{\pi} \right) \cdot \vec{\mathcal{T}} \, \chi_2 \  , \\
\mathcal{L}^{-}_{\pi \rho B_1 B_2} &= g_{\rho} \frac{f_{\pi B_1 B_2}}{m_{\pi}} \chi^{\dag}_1
\rho_{\mu} \left( \myatop{\mathbbm{1}}{0} \right)^{\mu}
\left( \vec{\rho} \times \vec{\pi} \right) \cdot \vec{\mathcal{T}} \, \chi_2 \  .
\end{align}
The purely mesonic interactions in Eqs.~(\ref{eq:WZ}) and (\ref{eq:RelLagrangians}) are unaffected by the $|\vec{p}_B|/m_B$ expansion since they have no dependence on baryon spinors.
The non-relativistic versions of Eqs.~(\ref{eq:omegaNN}) are
\begin{align}
\label{eq:omegaBB}
\mathcal{L}_{\omega B_1 B_2}^{+} &= g_{\omega B_1 B_2} \, \chi^{\dag}_1 \omega_{\mu}
\left( \myatop{\mathbbm{1}}{0} \right)^{\mu} \chi_2 \  ,
\nonumber \\
\mathcal{L}_{\omega B_1 B_2}^{-} &= g_{\omega B_1 B_2} \, \chi^{\dag}_1 \omega_{\mu}
\left( \myatop{0}{\vec{\mathcal{S}}} \right)^{\mu} \chi_2 \  .
\end{align}
In principle, these interactions require the same modifications as Eq.~(\ref{eq:rhoBB}) to satisfy a Ward-Takahashi identity.
However, in our analysis we will not be using the $\omega B_1 B_2$ vertex when the $\omega$ is an external particle,
as it is not part of a conserved vector current.  It then does not require the same modification as Eq.~(\ref{eq:rhoBB}) and
we may use Eq.~(\ref{eq:omegaBB}) without modification.

Before using this set of non-relativistic baryon propagators and Lagrangian interactions, together with the meson interactions,
to evaluate the Feynman diagrams shown in Figs.~\ref{fig:pi-cloud-cuts} and \ref{fig:omega-cloud-cuts}, we must account
for finite-size effects by implementing hadronic formfactors.

\subsection{Hadronic Formfactors}
\label{ssec:ff}
The above Lagrangian interactions treat particles as point-like objects. To account for finite-size hadronic effects, we use
the Pauli-Villars regularization via attaching a hadronic formfactor to each interaction vertex. However, maintaining gauge
invariance when implementing formfactors in scattering processes can be a challenging process.
For the $\pi B_1 \to \gamma B_2$ processes, we can insert formfactors on all $\pi B_1 B_2$ vertices in a
fully gauge-invariant manner. Formfactors for other interaction vertices will be handled in a more indirect manner.

\subsubsection{Gauge-Invariant Heavy-Pion Formalism}
\label{sssec_piB}
\begin{figure}[H]
\begin{center}
\subfloat[$t$-channel diagram 1]{
\includegraphics[scale=0.75]{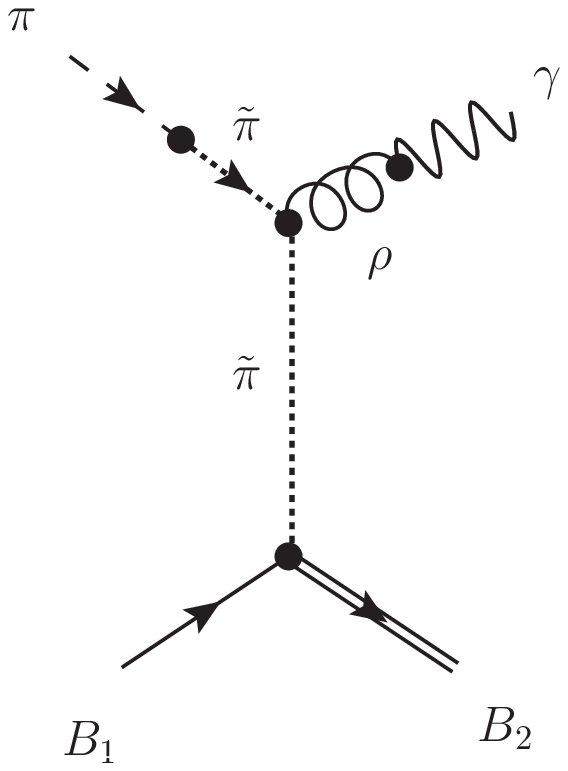}}
\subfloat[$t$-channel diagram 2]{
\includegraphics[scale=0.75]{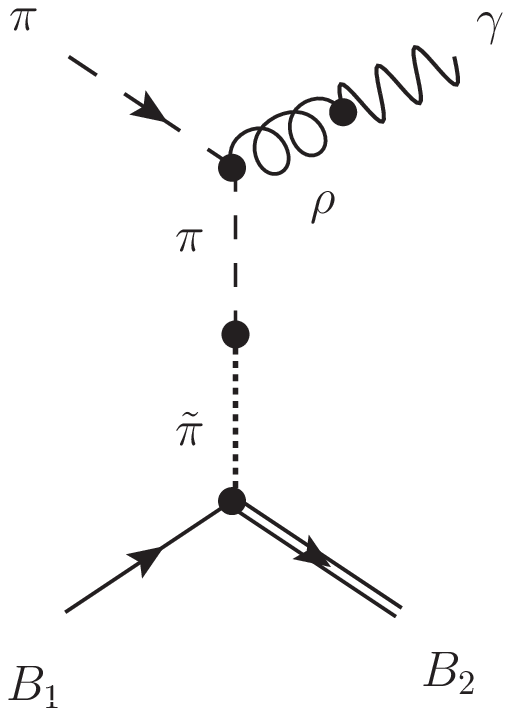}}
\end{center}
\caption{Two $t$-channel diagrams that amount to an implementation of $\pi B_1 B_2$ formfactors via application of Feynman rules.  The long dashed lines indicate ``normal'' pions while the short dashed lines indicate ``heavy'' pions.}
\label{fig:heavypions}
\end{figure}

In Ref.~\cite{Mathiot:1984bs} it was suggested that the insertion of a monopole $\pi NN$ formfactor, $\Lambda^2_{\pi}/(\Lambda^2_{\pi} + \vec{k}^2)$, in nucleon-nucleon scattering diagrams could be diagrammatically visualized as a fictitious ``particle" of mass $\Lambda_{\pi}$ with the same quantum numbers as the $\pi$-meson attaching to the ``normal'' pion lines,
see Figs.~7-9 in that work. Here, $\Lambda_{\pi}$ is the value of the formfactor cutoff and $\vec{k}$ is the pion's
three-momentum.  We shall denote this fictitious pion as $\tilde{\pi}$.  A rigorous way of using this ``heavy-pion'' method to
implement the $\pi NN$ and $\pi N \Delta$ formfactors was introduced in Refs.~\cite{Herrmann:1993za,Urban:1998eg}. There
it was shown that, by assigning appropriate Feynman rules for the inclusion of the heavy-pion propagator and $\pi\tilde\pi$
vertices, the resulting Feynman diagrams for $\rho$ selfenergies generate formfactors on all pertinent vertices thereby
maintaining gauge invariance. In Ref.~\cite{Roth} these Feynman rules were implemented in the context of $\pi N \to \rho N$
Born scattering diagrams, as opposed to selfenergies. There it was found that the gauge-invariant implementation of the
$\pi NN$ formfactor requires the inclusion of two $t$-channel terms: one where the heavy pion is attached to the external
pion line, and one where the heavy pion is attached to the internal pion line. These two diagrams are shown in
Fig.~\ref{fig:heavypions}. The remaining contact, $s$-, and $u$-channel diagrams have only the external pion to attach
to the heavy pion; therefore, the inclusion of the $\pi NN$ formfactor on those diagrams is straightforward.

As mentioned above, the application of these rules includes a $t$-channel diagram~\cite{Roth} containing a $\rho \tilde{\pi} \tilde{\pi}$ vertex term where the $\rho$ attaches to two heavy pions, shown in Fig.~\ref{fig:heavypions}. This introduces a
complication, as the structure of this vertex is \textit{a priori} not known. However, it was shown in Ref.~\cite{Urban:1998eg}
that a gauge-invariant heavy-pion vertex can be constructed by the requirement to satisfy the Ward-Takahashi identity,
\begin{equation}
\label{eq:heavypionWT}
q^{\mu} \tilde{\Gamma}_{\mu} \overset{!}{=} -g_{\rho} \left( G_{\tilde{\pi}}^{-1}(k-q) - G_{\tilde{\pi}}^{-1}(k) \right) \ ,
\end{equation}
where $q$ is the four-momentum of the incoming photon, $k$ is the four-momentum of the incoming heavy pion, and we
have suppressed isospin indices for simplicity.
Since the expression for the heavy-pion propagator is known, we can calculate the difference between the inverse
propagators:
\begin{align}
\label{eq:heavypiprop}
G_{\tilde{\pi}}^{-1}(k-q) - G_{\tilde{\pi}}^{-1}(k)  &=  \Lambda_{\pi}^2 + (\vec{k}-\vec{q}\,)^2
- \Lambda_{\pi}^2 -\vec{k}^2  \nonumber \,  \\
&= \vec{q} \cdot (\vec{q}-2\vec{k}) \, .
\end{align}
We then find that
\begin{equation}
\label{eq:heavyvertex}
q^{\mu} \tilde{\Gamma}_{\mu} \overset{!}{=}  -g_{\rho} \vec{q} \cdot (\vec{q}-2\vec{k}) \, .
\end{equation}
Since the four-momentum of the $\rho$ (which we denote by $q$) necessarily has a non-zero temporal component (as it
 is attached to a photon, so $q_0=|\vec{q}\,|$), the only possible $\rho$-heavy-pion interaction vertex which satisfies
the Ward-Takahashi identity is
\begin{equation}
\tilde{\Gamma}_{\mu} = -g_{\rho} \left( \myatop{0}{2\vec{k}-\vec{q}} \right)_{\mu} \, .
\label{eq:heavy-pi-rho-vertex}
\end{equation}
The monopole formfactor $\Lambda_{\pi}^2/(\Lambda_{\pi}^2 + \vec{k}^2)$ is sufficient to ensure convergence in the $S$-
and $P$-wave $\pi B_1 B_2$ interactions. However, our $D$-wave interaction, Eq.~(\ref{eq:Dwave}), contains two powers of
pion momentum so that the use of a monopole formfactor for this vertex does not generate sufficiently rapid convergence of the photon emission integral given by Eq.~(\ref{eq:rate}). Therefore, we introduce a dipole formfactor of the form
$\Lambda_{\pi}^4/(\Lambda_{\pi}^4 + \vec{k}^4)$. We may then introduce the formfactor in a gauge-invariant manner using
the following  Feynman rules~\cite{Urban:1998eg} for processes involving a $D$-wave $\pi B_1 B_2$ interaction:
\begin{enumerate}
\item A heavy pion ($\tilde{\pi}$) attaches to a normal pion in all possible places,
\item the heavy-pion ``propagator'', $G_{\tilde{\pi}}$, receives a factor of $-i / (\Lambda^4_{\pi} + \vec{k}^4)$, and
\item the pion--heavy-pion vertex receives a factor of $i\Lambda_{\pi}^4$.
\end{enumerate}
Implementing these rules gives the same results as for the monopole formfactor, but with higher powers of $\Lambda_{\pi}$
and $\vec{k}$. However, since we have altered the propagator of the heavy pion, the above result for the
$\rho \tilde{\pi} \tilde{\pi}$ vertex, Eq.~(\ref{eq:heavyvertex}), no longer applies. We must repeat the procedure using the
new $1/(\Lambda_{\pi}^4 + \vec{k}^4)$ propagator to construct a new $\rho \tilde{\pi} \tilde{\pi}$ vertex. The structure of
the Ward-Takahashi identity is identical, but the difference in the inverse heavy-pion propagators becomes
\begin{align}
\label{eq:Dwavepropdiff}
G_{\tilde{\pi}}^{-1}(k-q) - G_{\tilde{\pi}}^{-1}(k)  &= \Lambda_{\pi}^4 + (\vec{k}-\vec{q}\,)^4
- \Lambda_{\pi}^4 -\vec{k}^4  \nonumber \, , \\
&=   - 4\vec{k}^2 (\vec{k}\cdot\vec{q}) + 4(\vec{k}\cdot\vec{q}\,)^2
+2\vec{k}^2 \vec{q}^{\, 2} - 4(\vec{k}\cdot\vec{q}) \vec{q}^{\,2} +\vec{q}^{\,4} \nonumber \, \\
&= \left[ -4\vec{k}^2 \vec{k} + 4(\vec{k}\cdot\vec{q}\,)\vec{k} +2\vec{k}^2 \vec{q} + \vec{q}^{\,2} \vec{q} \,
\right] \cdot\vec{q} - 4(\vec{k}\cdot\vec{q}\,) \vec{q}^{\,2} \ .
\end{align}
Here we encounter an ambiguity we did not have with the monopole formfactor. We need to ``factor out'' a
$\vec{q}$ from the above expression in order to identify the vertex, which we have already done to the
term in brackets. In Eq.~(\ref{eq:heavy-pi-rho-vertex}) there is one unique choice for this. However, in the second term in
Eq.~(\ref{eq:Dwavepropdiff}) we may factor out the $\vec{q}$  from either the $(\vec{k}\cdot\vec{q}\,)$ or the
$\vec{q}^{\,2}$ term.  We have checked both choices and found that there is a negligible difference between the resulting
photon emission rates for all photon energies. Therefore, for our purpose of calculating thermal photon emission rates
either choice is fine; we choose to factor out the $\vec{q}$ from the $\vec{q}^{\,2}$ term in
Eq.~(\ref{eq:Dwavepropdiff}). This leads to a $D$-wave $\rho \tilde{\pi} \tilde{\pi}$ vertex of the form
\begin{equation}
\tilde{\Gamma}_{\mu}^D = -g_{\rho} \left( \myatop{0}{ 4\vec{k} (\vec{k}^2 - \vec{k}\cdot\vec{q} \, )
+ \vec{q} \,(4\vec{k} \cdot \vec{q} -2\vec{k}^2-\vec{q}^{\, 2}) } \right)_{\mu} \, .
\end{equation}

\subsubsection{Gauge-Invariant Meson Formfactors}
\label{sssec_piroom}
We now have defined our $\pi B_1 B_2$ formfactors and established an implementation that ensures gauge invariance.
However, there remain vertices in the $\pi B_1 \to \gamma B_2$ processes which do not have formfactors applied to them. For
example, the $s$-channel diagram shown in Fig.~\ref{fig:pi-cloud-cuts}(a) our above method applies a formfactor to the
$\pi B_1 B_2$ vertex on the left of the diagram. However, we have not applied a formfactor to the $\rho B_2 B_2$ vertex.
Similarly, in the $t$-channel diagram of Fig.~\ref{fig:pi-cloud-cuts}(b), the $\rho \pi \pi$ vertex at the top of the diagram also
lacks a formfactor. In order to fully account for finite-size effects, we employ the method used in
Refs.~\cite{Turbide:2003si,Holt:2015cda}. We identify the dominant diagram, which is the $t$-channel pion exchange.
We then apply a factorized formfactor using an average pion exchange momentum. Then we apply a dipole $\rho \pi \pi$
formfactor  to the overall scattering  process~\cite{Rapp:1999qu},
\begin{equation}
FF_{\rho \pi\pi}(\bar{t}\,) =\left(  \frac{2\Lambda_{\rho \pi\pi}^2}{2\Lambda_{\rho \pi\pi}^2 - \bar{t}}\right)^2 \ ,
\end{equation}
where we evaluate the average exchange momentum, $\bar{t}$, via the expression
\begin{align}
\left( \frac{1}{m_{\pi}^2 - \bar{t}} \right)^2 & \left( \frac{2\Lambda_{\rho \pi\pi}^2}{2\Lambda_{\rho \pi\pi}^2 - \bar{t}} \right)^4 \nonumber \\ &= - \frac{1}{4 q_0^2} \int_{0}^{-4 q_0^2} dt \left( \frac{1}{m_{\pi}^2 - t} \right)^2 \left( \frac{2\Lambda_{\rho \pi\pi}^2}{2\Lambda_{\rho \pi\pi}^2 -t} \right)^4 \ .
\label{eq:tbar2}
\end{align}
We use $\Lambda_{\rho \pi\pi}=1$ GeV in accordance with Refs.~\cite{Rapp:1999qu,Turbide:2003si,Holt:2015cda}.
This averaged formfactor is then applied to the overall amplitude appearing in Eq.~(\ref{eq:rate}):
\begin{equation}
\overline{|M|^2} \to \overline{|M|^2}\, FF(\bar{t})^4 \, .
\end{equation}
This method accounts for formfactor effects that are not incorporated with the heavy-pion technique.  This is the final ingredient
for formfactor implementation in $\pi B_1 \to \gamma B_2$ processes.

\subsubsection{$\omega B_1 B_2$ Processes}
\label{sssec_omB}
Due to the relatively large $\pi \rho \omega$ coupling, we consider two other processes involving the $\omega$ meson,
shown previously in Fig.~\ref{fig:omega-cloud-cuts}. The process shown in Fig.~\ref{fig:omega-cloud-cuts}(a) involves
the $\omega B_1 B_2$ vertex of Eq.~(\ref{eq:omegaBB}), where the $\omega$ is an exchange particle. For this vertex,
we use the standard monopole formfactor $\Lambda^2 / (\Lambda^2 + \vec{k}^2)$ with $\vec{k}$ being the momentum
of the $\omega$. The second process, shown in Fig.~\ref{fig:omega-cloud-cuts}(b), involves the $\omega$ as an external
particle, attaching to the $\pi \rho\omega$ vertex of Eq.~(\ref{eq:WZ}). As this is a purely mesonic vertex, we use a dipole
formfactor~\cite{Rapp:1999qu} of $\left[ 2\Lambda^2 / (2\Lambda^2-t^2) \right]^2$.  Due to the gauge invariance of
the $\pi \rho \omega$ photo-emission vertex in both processes, we only need to consider the $t$-channel diagrams. This
allows a straightforward implementation of formfactors on both vertices without the need to resort to a factorized formfactor.

\subsection{Evaluation of Parameters}
\label{ssec:para}
The large number of vertices involved in our photoemission processes leaves us with a similarly large number of parameters.
The purely mesonic parameters we will be using have already been evaluated in Ref.~\cite{Holt:2015cda}. In addition,
the following quantities must be evaluated:
\begin{itemize}
\item the coupling constant $f_{\pi B_1 B_2}$ for each possible ${\pi B_1 B_2}$ vertex, where $B_1$ and $B_2$ are any of the baryons under consideration,
\item the cutoff $\Lambda_{\pi B_1 B_2}$ for each $\pi B_1 B_2$ vertex formfactor,
\item the coupling constant $g_{\omega B_1 B_2}$ for each possible ${\omega B_1 B_2}$ vertex, and
\item the cutoff $\Lambda_{\omega B_1 B_2}$ for each $\omega B_1 B_2$ vertex formfactor.
\end{itemize}

\subsubsection{Coupling Constants}
\label{sssec_cc}
We use data from the Particle Data Group~\cite{Olive:2016xmw} on $B_1 \to \pi B_2$ decays to calculate the
$f_{\pi B_1 B_2}$ coupling constants. Only ``established'' baryons, \ie,  nucleons, deltas, and hyperons with a
three- or four-star status, are included. The data used to calculate decays are collected in the Appendix~\ref{sec:app}.
We neglect all $\pi B_1 \to \gamma B_2$ processes which contain couplings that cannot be calculated due to lack of
available decay data. The coupling constants are found by applying Feynman rules to $B_1 \to \pi B_2$ decay processes
and inserting the resulting amplitude into the standard two-particle decay formula. In the rest frame of $B_1$, the latter
is given by~\cite{Halzen:1984mc}
\begin{equation}
\label{eq:decayformula}
\Gamma_{B_1 \to \pi B_2} = \frac{{p}_{CM}}{8 \pi m_{B_1}^2} \overline{|M|^2} \, FF({p}_{CM})^2 \, ,
\end{equation}
where $\Gamma_{B_1 \to \pi B_2}$ is the partial width for the decay process, $p_{CM}$ is the magnitude of the center-of-mass
three-momentum of each daughter particle, $FF(p_{CM})$ is the formfactor for the $\pi B_1 B_2$ vertex, and $\overline{|M|^2}$
denotes the initial-state averaged and final-state summed squared matrix element. This amplitude contains the (squared) coupling
we wish to evaluate. The center-of-mass momentum can be calculated straightforwardly by applying conservation of
four-momentum to the invariant mass of the parent particle.
Many resonances more massive than the $\Delta(1232)$ have considerable uncertainty in both their total widths and
branching ratios, both of which are needed to evaluate the $\pi B_1 B_2$ couplings. To account for this uncertainty, we
introduce an uncertainty parameter $0.6 \le y \le 1.4$ which multiplies the partial width in Eq.~(\ref{eq:decayformula}), so
that $\Gamma_{B_1 \to \pi B_2} \to y \Gamma_{B_1 \to \pi B_2}$.  Additionally, to simplify calculations, we use the same
formfactor cutoff for all resonances other than the nucleon and $\Delta(1232)$, so that we have only three cutoffs for
$\pi B_1 B_2$ interactions:
$\Lambda_{\pi NN}$, $\Lambda_{\pi N \Delta}$, and $\Lambda_{\pi BB}$.

This method of calculating couplings is appropriate when the decay products have zero or relatively small width. However, if the daughter baryon has a non-negligible width, treating it as a stable particle may no longer be justified, especially if here is limited decay phase space (\ie, when the sum of the daughter masses is close to the parent mass).  We can then no longer
use a fixed-mass approximation for the daughter baryon's spectral distribution, and rather should integrate over its invariant
mass.  The pertinent extension of Eq.~(\ref{eq:decayformula}) reads~\cite{Rapp:1997fs}
\begin{equation}
\Gamma_{B_1 \to \pi B_2} = \int dm \, m \, \frac{p_{CM}(m)}{8 \pi m_{B_1}^2} \,
\overline{|M(q)|^2} \, FF(p_{CM}(m))^2 \, \rho_{B_2}(m) \ ,
\label{eq:decayformula2}
\end{equation}
where $m $ is the (variable) mass of the daughter baryon $B_2$, and $\rho_{B_2}(m)$ is its SF.  For simplicity, we model $\rho_{B_2}(m)$ using a Breit-Wigner resonance with a mass-dependent width,
\begin{equation}
\rho_{B_2}(m) = -\frac{1}{\pi} \mathrm{Im} \, D_{B_2}(m) = \frac{1}{\pi}
\frac{m \Gamma(m)}{\left(m^2 -m_{B_2}^2\right)^2 + m^2 \Gamma(m)^2} \ .
\end{equation}
The width of $B_2$ is generated by the decay process $B_2 \to \pi N$.  This gives us the lower bound for the integration
over $m$ in Eq.~(\ref{eq:decayformula2}) as $m_{\rm min} = m_N+m_{\pi}$. The upper bound for $q$ is given by the
amount of energy available for $B_2$, $m_{\rm max}=m_{B_1}-m_{\pi}$. These limits also give us a criterion for when
we need to use Eq.~(\ref{eq:decayformula2}) over Eq.~(\ref{eq:decayformula}).  Quantitatively, we integrate over the
SF when particle masses and widths satisfy the condition
\begin{equation}
m_{B_1} - (m_{B_2}+m_{\pi}) < \frac{\Gamma_{B_2}}{2}  \, .
\end{equation}
In practice, only one of our couplings requires this treatment, namely for the $\Delta(1600) \to \pi N(1440)$ decay.
The $N(1440)$ has a width of 350 MeV, and $m_{\Delta(1600)} - (m_{N(1440)}+m_{\pi}) = 20$ MeV, which necessitates
the use of Eq.~(\ref{eq:decayformula2}). The calculation of the $\pi N(1440)\Delta(1600)$ coupling with a sharp
$N(1440)$ mass using Eq.~(\ref{eq:decayformula}) gives $f_{\pi N(1440) \Delta(1600)} = 8.4$, which turns out to
be a gross overestimate. The calculation using Eq.~(\ref{eq:decayformula2}) and a finite $N(1440)$ width gives
$f_{\pi N(1440) \Delta(1600)} = 4.9$.

We may calculate the $g_{\omega B_1 B_2}$ couplings in the same manner as the $f_{\pi B_1 B_2}$ couplings. Since
the $\omega$ is an isoscalar particle it can couple only to baryons with identical isospin; there are no $\omega N \Delta$
couplings.  Additionally, since the $\omega$ is not easily reconstructed from its dominant 3$\pi$ decay, direct data on
$B_1 \to \omega B_2$ decays is greatly lacking. However, it was found in Ref.~\cite{Post:2000rf} that one can use
helicity amplitudes of $N^{*} \to \gamma N$ decays together with the vector meson dominance model to indirectly
estimate $\omega N N^{*}$ couplings. This also allows us to calculate couplings that occur below the $\omega$
production threshold. The couplings are found by equating Eq.~(\ref{eq:decayformula}) with the expression for the
partial width of a radiative decay in terms of helicity amplitudes, given by~\cite{Olive:2016xmw}
\begin{equation}
\Gamma_{N^* \to \gamma N} = \frac{\vec{p}_{CM}^{\,2}}{\pi} \frac{2m_N}{(2J+1) m_{N^{*}}} \left( |A_{1/2}|^2 + |A_{3/2}|^2 \right) \, ,
\end{equation}
where $J$ is the spin of the parent particle. We note that $A_{3/2} = 0$ for radiative decays of spin-1/2 resonances. It was
shown in Ref.~\cite{Post:2000rf} that by taking the appropriate combinations of proton and neutron helicity amplitudes, one
can isolate the contributions from the isoscalar ($\omega$) and isovector ($\rho$) channels. These combinations are
\begin{equation}
A_i^s = \frac{1}{2} (A_i^p + A_i^n), \quad A_i^v = \frac{1}{2} (A_i^p - A_i^n) \, ,
\end{equation}
where $s$ and $v$ indicate the isoscalar and isovector combinations, respectively, and $i$ is $1/2$ or $3/2$. Using the
isoscalar combination of helicity amplitudes then allows us to solve for $g_{\omega N N^*}$.

The final step in quantifying the $\pi B_1 B_2$ and $\omega N N^*$ couplings requires to establish values for the formfactor
cutoffs, $\Lambda_{\pi NN}$, $\Lambda_{\pi N \Delta}$, $\Lambda_{\pi BB}$, and $\Lambda_{\omega NN^*}$. Information
on the $\pi B_1 B_2$ cutoffs can be inferred by fitting phase shift data for elastic $\pi N$ scattering. However, there are
several constraints on parameter choices we must observe.

\subsubsection{Formfactor Cutoffs}
\label{sssec_photoabs}
\begin{figure}[H]
\begin{center}
\includegraphics[scale=1.0]{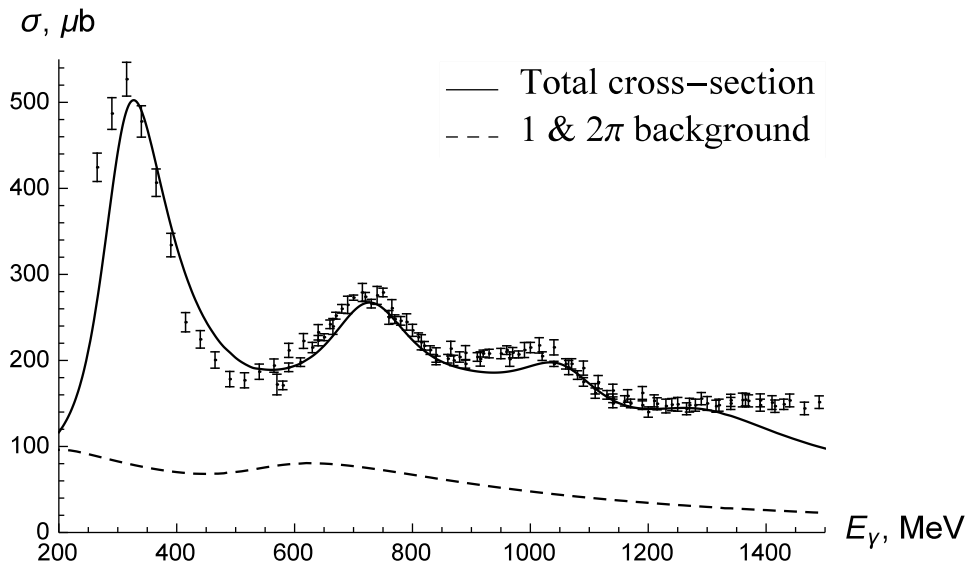}
\end{center}
\caption{Calculation of the photoabsorption cross section on the proton with $\Lambda_{\pi NN}=310$ MeV (solid line),
and 1- and 2-$\pi$ production background (dashed line) from Refs.~\cite{Rapp:1997fs,Rapp:1999ej}. The experimental
data are from Refs.~\cite{Armstrong:1971ns,Bartalini:2008zza}.}
\label{fig:photoabsorption}
\end{figure}

Previous calculations of the in-medium $\rho$ spectral
function~\cite{Rapp:1997fs,Urban:1998eg,Urban:1999im,Rapp:1999us,Rapp:1999qu}, which serve as our benchmark,
found that the $\pi NN$ formfactor cutoff, $\Lambda_{\pi NN}$, could be no larger than $\approx 500$~MeV in order to
remain consistent with photoabsorption data on the proton, see Fig.~\ref{fig:photoabsorption}. Larger values yielded
non-resonant background cross sections that
exceeded experimental data.  Furthermore, a calculation of the $\pi^{-} p \to \rho^0 n$ cross section using the same
vertices employed here~\cite{Roth} found that consistency with experimental data demanded that $\Lambda_{\pi NN}$
be $\approx 310$ MeV using a coupling of $f_{\pi NN}=1$. We therefore take this cutoff value and coupling as restricted,
allowing only a possible 10\% variation of their values.

A second constraint applies to the $\pi N \Delta$ formfactor cutoff.  The works we are augmenting use a cutoff of
$\Lambda_{\pi N \Delta} = \Lambda_{\pi NN} = 310$~MeV. This value is constrained by the 2$\pi$ production
contribution to the total proton photoabsorption cross section.  The pioneering works which evaluated $\Lambda_{\pi N \Delta}$
via fits to $P_{33}$ phase shift data found excellent agreement with a value of
360~MeV~\cite{Lenz:1975qx,Woloshyn:1976ca,Moniz:1981zz}. We therefore allow our cutoff to vary up to this value.
Additionally, we allow the values of $\Lambda_{\pi BB}$ to vary from 310 MeV (to match the $\pi NN$ cutoff) up to a
value of 1500 MeV, which is a typical size for formfactor cutoffs in the Bonn nucleon-nucleon potential
model~\cite{Machleidt:1989tm}.

We are also constrained in our choices for the $\omega NN$ coupling and formfactor. The process $\gamma N \to \pi N$ via
$\omega$ $t$-channel exchange contributes to the photoabsorption cross section on the proton. Since this process was not
included in the fits using the in-medium $\rho$ SF~\cite{Rapp:1997fs,Rapp:1999ej}, we add the
$\omega$ $t$-channel photoabsorption cross
section to the overall result. Our choice of coupling and formfactor should not raise the total cross section above the
experimental data. The resonance couplings of $\omega N N^*$ are similarly constrained by the cross sections of
$\gamma N \to \omega N^*$ photoproduction processes.

\subsubsection{$\pi N$ Scattering Phase Shifts and Summary of Parameter Values}
\label{sssec_piN}
To evaluate the parameters $f_{\pi NN}$, $\Lambda_{\pi N \Delta}$, and $\Lambda_{\pi BB}$, we fit phase shift data
for elastic $\pi N \to \pi N$ scattering in the $P_{11}$ and $P_{33}$ channels. We neglect the $S$-wave channel since
these involve $t$-channel diagrams with the exchange of $\rho$ mesons. These diagrams include a $\rho BB$ interaction.
Since we are not using the $\rho BB$ formfactor in our photoemission rate calculations, calculation of the $S$-wave phase
shifts would involve introducing an extra parameter, $\Lambda_{\rho BB}$, that would not enter into our final calculations
for photon rates. The $P_{13}$ and $P_{31}$ channels are neglected due to the relatively small size of their phase shift,
$\delta \lesssim 5^{\circ}$.

\begin{figure}
\subfloat[$P_{11}$ channel]{
\includegraphics[scale=0.65]{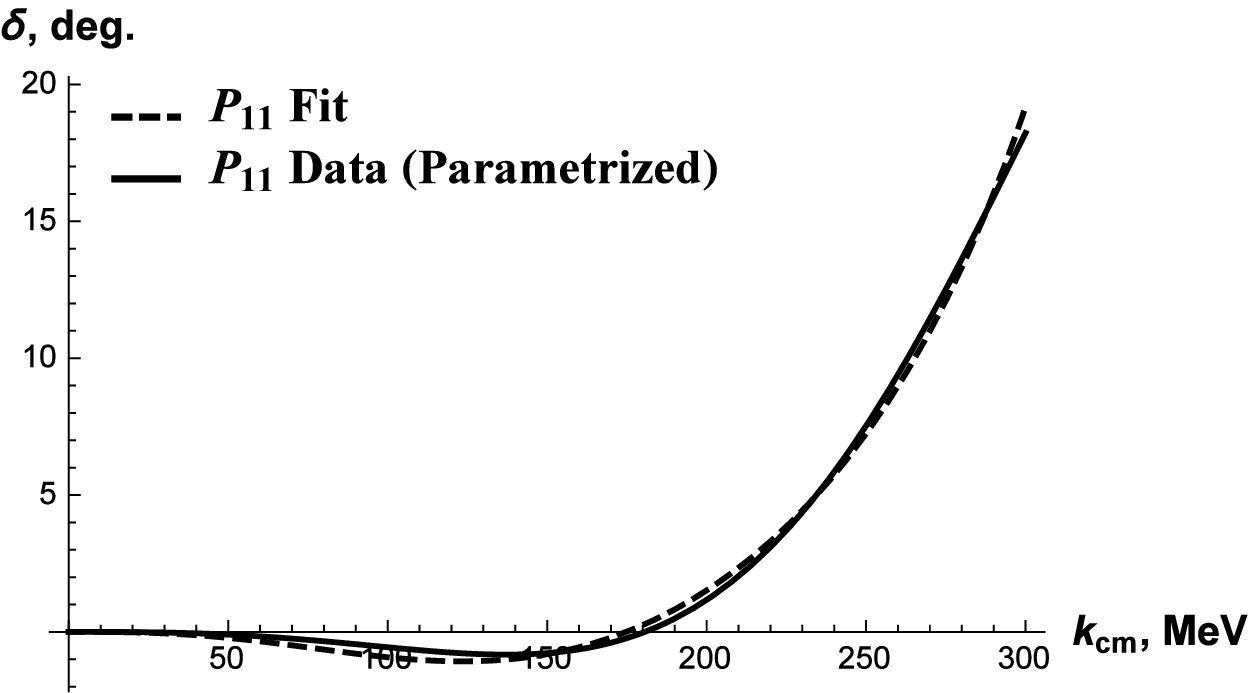}}
\subfloat[$P_{33}$ channel]{
\includegraphics[scale=0.65]{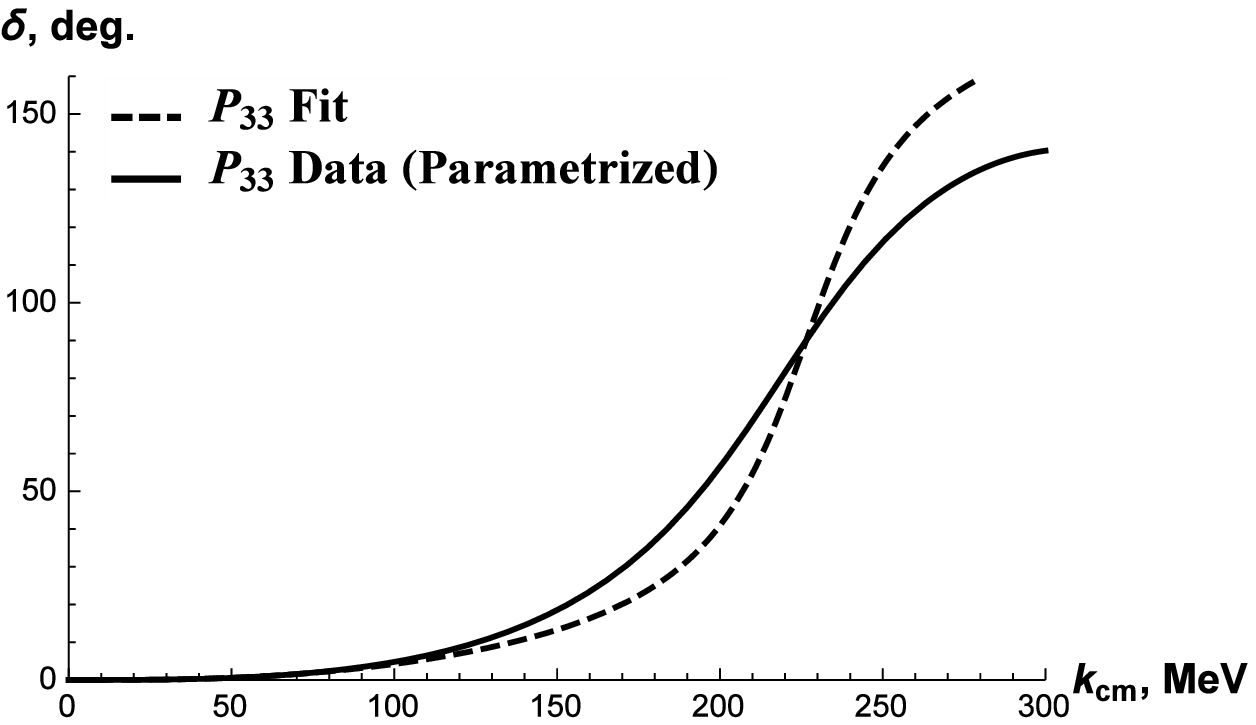}}

\subfloat[$P_{11}$ channel]{
\includegraphics[scale=0.65]{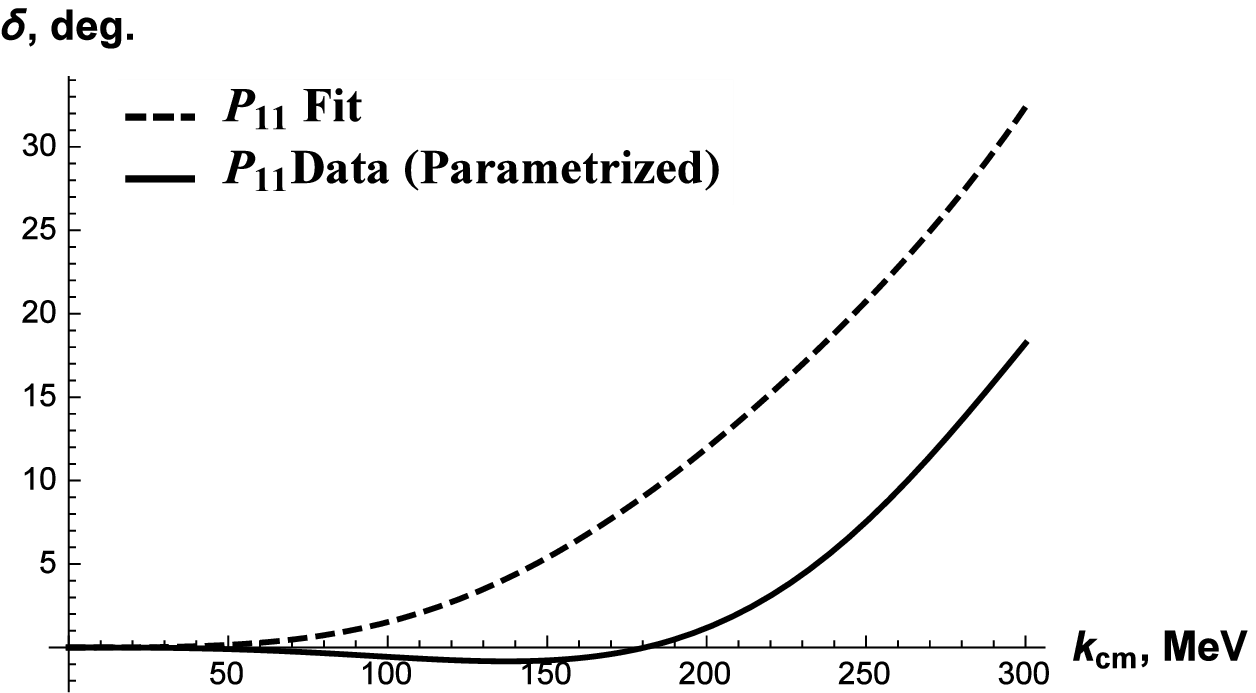}}
\subfloat[$P_{33}$ channel]{
\includegraphics[scale=0.65]{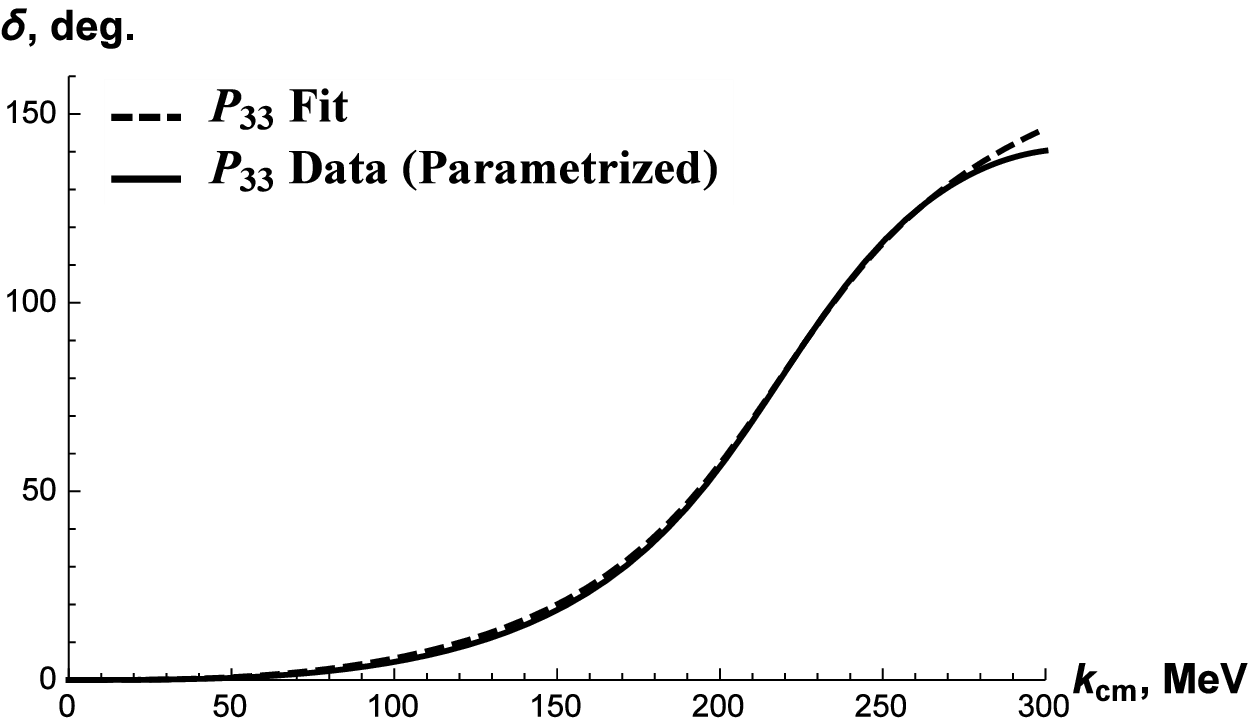}}
\caption{Pion nucleon scattering phase shifts with parameters fit to $P_{11}$ data only (top row) and with parameters fit
to $P_{33}$ data only (bottom row).}
\label{fig:P11and33fit}
\end{figure}

The $P$-wave phase shifts are relatively easily calculated using the $K$-matrix formalism~\cite{Chung:1995dx}. For elastic
scattering, the relation between the $K$-matrix and the phase shift is $K_{\alpha} = \frac{1}{k} \tan{\delta_{\alpha}}$,
where $k$ is the magnitude of the center-of-mass three-momentum and $\alpha$ is a given partial wave and isospin channel.
The relativistically improved $K$-matrix (RIKM) model~\cite{Oset:1981ih,Ericson:1988gk} provides a suitable way
to carry this out. It is composed of Born diagrams using non-relativistic interactions identical to ours.  Energy denominators
for the $s$- and $u$-channels are then treated relativistically as shown below. This relativistic treatment also involves
moving beyond the static approximation where nucleon momenta are neglected, \ie, center-of-mass momentum is used
instead of simply the momentum of the incoming pion.

\begin{table}
\begin{center}
\begin{tabular}{|c|c|c|c|c|}
\hline
& \begin{tabular}{@{}c@{}} $P_{11}$ Only \\ Fit \end{tabular} & \begin{tabular}{@{}c@{}} $P_{33}$ Only\\ Fit \end{tabular} & \begin{tabular}{@{}c@{}}Simultaneous \\ Fit \end{tabular} & \begin{tabular}{@{}c@{}}Weighted \\ Fit \end{tabular} \\
\hline
$f_{\pi NN}$ & 1.1 & 0.9 & 1.1 & 1.1 \\
\hline
$y$ & 0.69 & 0.6 & 0.6 & 0.6 \\
\hline
$\Lambda_{\pi N \Delta}$ & 310 & 360 & 360 & 360 \\
\hline
$\Lambda_{\pi BB}$ & 1360 & 410 & 520 & 920 \\
\hline
\end{tabular}
\end{center}
\caption{Parameter combinations for $\pi N$ partial-wave phase shift fits.}
\label{tab:parameters}
\end{table}

As an example, the contributions to $K_{11}$ from the nucleon and $\Delta(1232)$ are given by
\begin{align}
K_{N} &= \frac{1}{4\pi} \frac{m_N}{\sqrt{s}} \frac{2 m_N \vec{k}^2}{3} \frac{f_{\pi NN}^2}{m_{\pi}^2}
\left( \frac{\Lambda_{\pi NN}^2}{\Lambda_{\pi NN}^2 +\vec{k}^2} \right)^2
\left( \frac{9}{s-m_N^2}+ \frac{1}{u-m_N^2} \right) \, , \nonumber \\
K_{\Delta} &= \frac{1}{4\pi} \frac{m_N}{\sqrt{s}} \frac{2 m_{\Delta} \vec{k}^2}{3} \frac{f_{\pi N \Delta}^2}{m_{\pi}^2}
\left( \frac{\Lambda_{\pi N \Delta}^2}{\Lambda_{\pi N \Delta}^2 +\vec{k}^2} \right)^2
\frac{16}{9} \frac{1}{u-m_{\Delta}^2} \, ,
\end{align}
where $s=(\omega_N + \omega_{\pi})^2$, $\omega_N = \sqrt{m_N^2 + \vec{k}^2}$,
$\omega_{\pi} = \sqrt{m_{\pi}^2 + \vec{k}^2}$, and $u$ is approximated by
$u \approx m_B^2 +m_{\pi}^2 +2 \omega_N \omega_{\pi}$~\cite{Ericson:1988gk}.
Contributions from other resonances and for other channels follow by constructing the relevant Born diagrams and evaluating
the projections into the pertinent spin and isospin channel.
We include all $P$-wave nucleon and delta resonances up to a mass of 1.9\,GeV in our list of particles (see
Appendix~\ref{sec:app}), namely the $N(1440)$, $N(1710)$, $N(1720)$, $N(1900)$, $\Delta(1232)$, $\Delta(1600)$, and
$\Delta(1910)$.

\begin{figure}[H]
\begin{center}
\subfloat[$P_{11}$ channel]{
\includegraphics[scale=0.65]{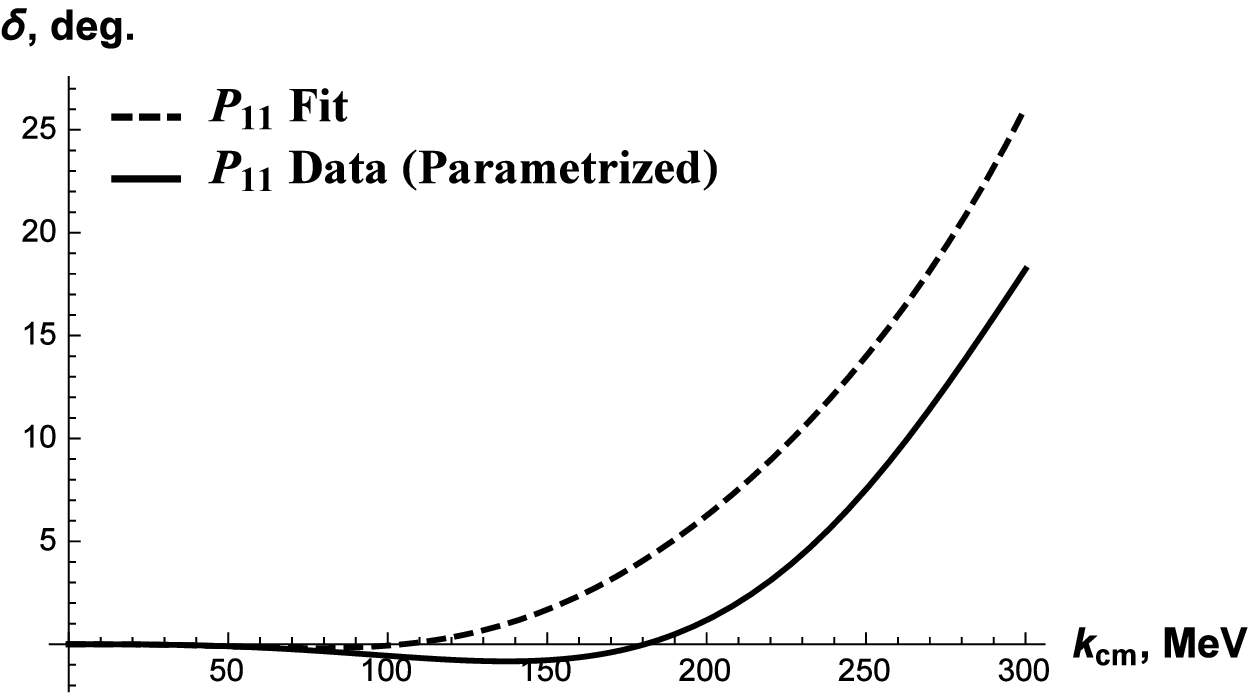}}
\subfloat[$P_{33}$ channel]{
\includegraphics[scale=0.65]{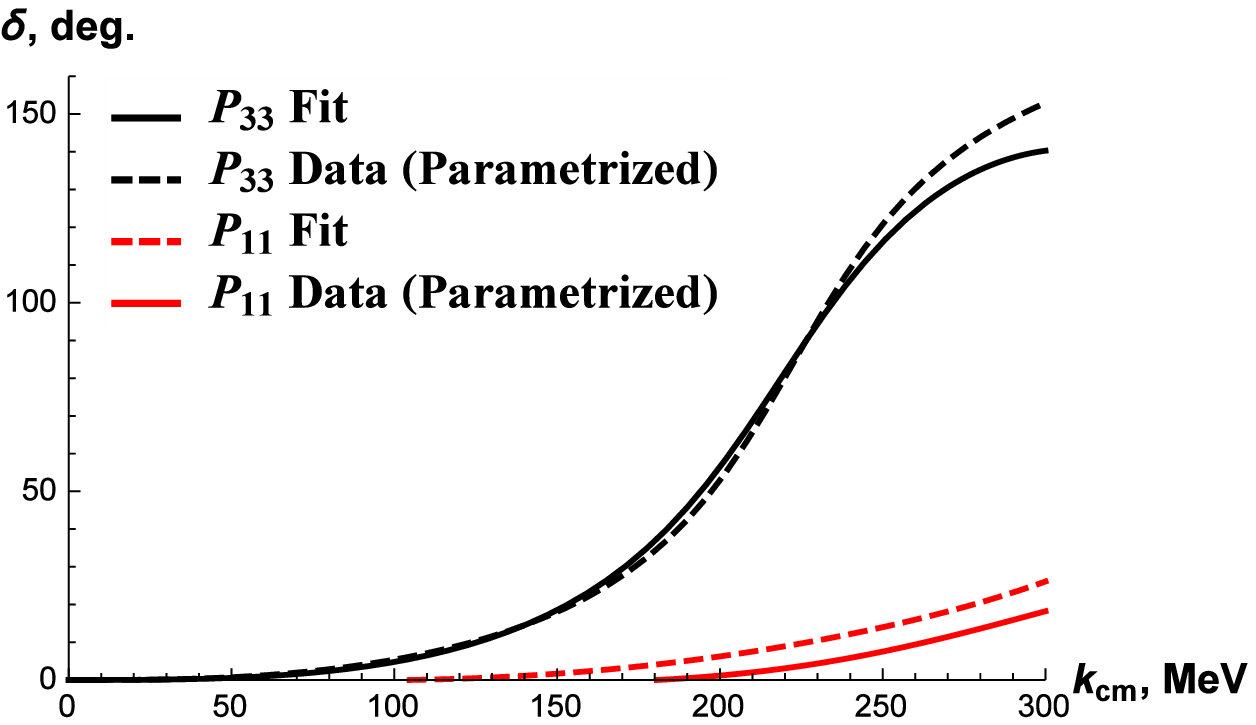}}

\subfloat[$P_{11}$ channel]{
\includegraphics[scale=0.65]{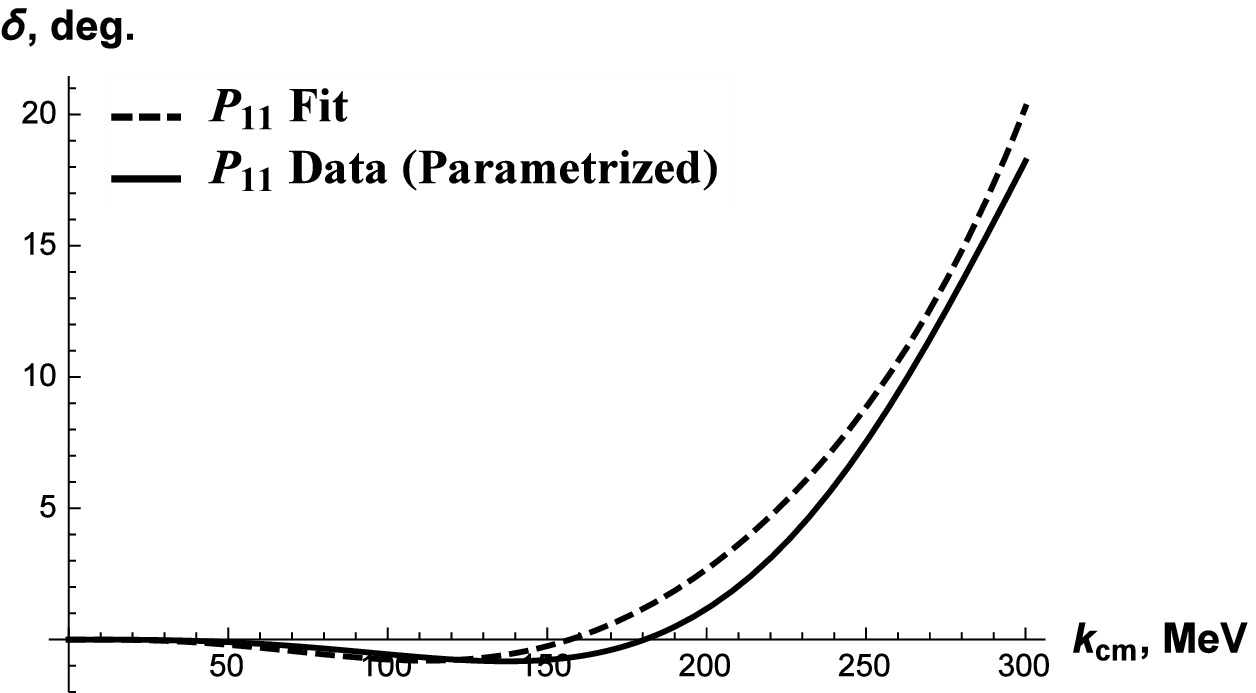}}
\subfloat[$P_{33}$ channel]{
\includegraphics[scale=0.65]{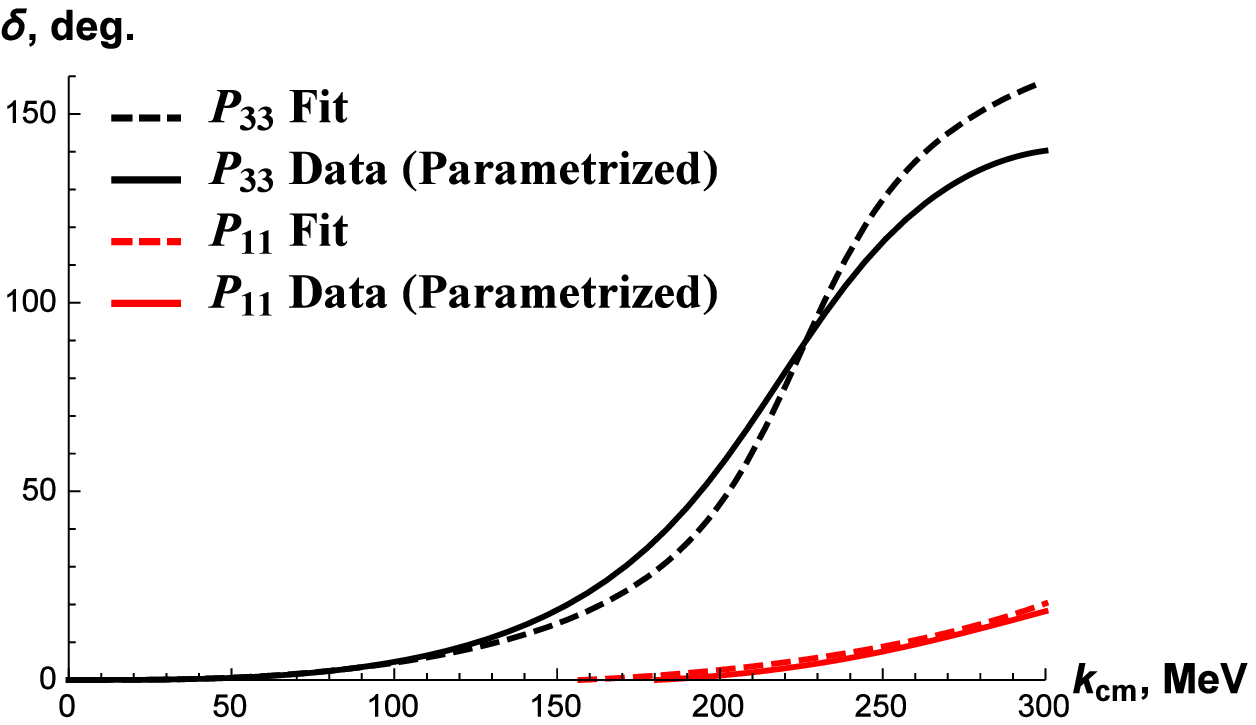}}
\end{center}
\caption{Pion-nucleon scattering phase shifts with parameters simultaneously fit to both channels (top row), and with
parameters fit to both channels with increased weighting toward the $P_{11}$ channel (bottom row).}
\label{fig:bothfit}
\end{figure}

We now use the RIKM model to evaluate our parameters $\Lambda_{\pi N \Delta}$, $\Lambda_{\pi BB}$, and $f_{\pi NN}$
by fitting phase shifts to the data from Ref.~\cite{Workman:2012hx} for center-of-mass momenta from 0 to 300 MeV.
In principle, we should limit our analysis to the $\pi$ production threshold of $k_{\mathrm{cm}}\approx 215$ MeV.  After this
point the phase shift acquires an imaginary part, indicating the onset of inelasticity in the scattering channel. However, we
have verified that there is a negligible difference in the resulting parameter fits when fitting phase shifts up to 215 MeV
versus a maximum value of 300 MeV. We note that the phase shift is much larger in the $P_{33}$ channel than in the
$P_{11}$ channel.

The fits are performed by minimizing the integrated difference between our $K$-matrix phase shift and the data fit
from Ref.~\cite{Workman:2012hx}. We define this difference to be
\begin{equation}
\label{eq:fit}
D = \int_0^{k_{\mathrm{max}}} dk \left\{ \left[ \delta_{11}^{\mathrm{data}}(k)- \delta_{11}^{\mathrm{RIKM}}(k) \right]^2 +
\left[ \delta_{33}^{\mathrm{data}}(k)- \delta_{33}^{\mathrm{RIKM}}(k)\right]^2 \right\} \, ,
\end{equation}
where $\delta^{\mathrm{data}}$ is the fit to data from Ref.~\cite{Workman:2012hx} and $\delta^{\mathrm{RIKM}}$ is our phase shift calculated
using the RIKM model.  In Fig.~\ref{fig:P11and33fit} we display the results for the fits that result from optimizing the parameters
for one channel at a time. We see that the parameter combination which provides an optimal fit in one channel does not yield
good a fit in the other channel. This suggests that we need to find a suitable balance of parameter values which better satisfies
both channels. We show the simultaneous fit of both channels given by Eq.~(\ref{eq:fit}) in Figs.~\ref{fig:bothfit}(a) and (b).
Since the $P_{33}$ channel phase shift varies from 0 to $\approx 150^{\circ}$ while the $P_{11}$ channel phase shift varies
from 0 to $\approx 20^{\circ}$, the former carries a larger weight and it appears that the $P_{11}$ channel has a worse fit
than the $P_{33}$ channel; in particular, this fit does not display the repulsive (negative) phase shift in the $P_{11}$ channel
at momenta below $\approx 180$ MeV which is featured in the data.  We can remedy this issue by giving a greater weight to
the $P_{11}$ channel in Eq.~(\ref{eq:fit}); with an extra weight factor of 10 of the  $P_{11}$ relative to the $P_{33}$
channel, the only parameter that changes is $\Lambda_{\pi B_1 B_2}$, increasing from 520 MeV to 920 MeV.
These fit results, shown in Figs.~\ref{fig:P11and33fit}(c) and (d), now display the desired repulsion in the $P_{11}$ channel
at the cost of a slightly worse fit in the $P_{33}$ channel. We shall use the parameters from the fit shown in
Fig.~\ref{fig:bothfit}(c) and (d) for our calculations, \ie,  $\Lambda_{\pi B_1 B_2} = 920$ MeV.  However, since couplings
and formfactors figure into the decay rates as $f^2 FF^2$, changing (increasing) the formfactor cutoff causes a
compensatory (decreasing) effect on the couplings. We have found that the resulting seesaw effect produces a negligible
difference in our photon emission rates when using a value of $\Lambda_{\pi B_1 B_2} = 520$ MeV.

\begin{figure}[H]
\begin{center}
\includegraphics[scale=1.0]{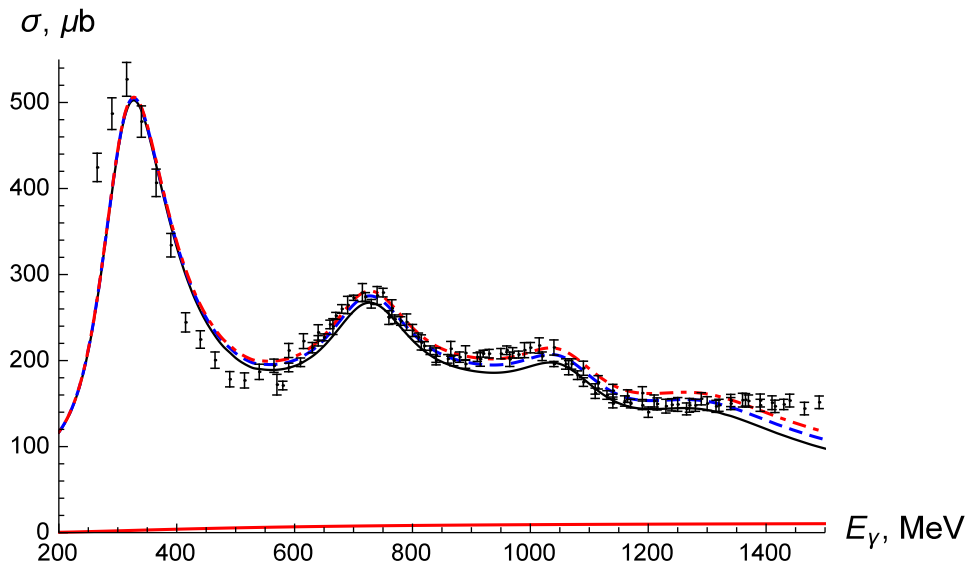}
\end{center}
\caption{Calculations of the total photoabsorption cross section on the proton without $\omega$ $t$-channel exchange (solid
black line), and including the $t$-channel exchange with $\Lambda_{\omega NN}=500$ MeV (dashed blue line) and with
$\Lambda_{\omega NN}=750$ MeV (dot-dashed red line). The lower solid red line is the individual  $\omega$ $t$-channel
contribution with $\Lambda_{\omega NN}=500$ MeV. Data are from Refs.~\cite{Armstrong:1971ns,Bartalini:2008zza}.}
\label{fig:photoabsorption-omega}
\end{figure}

To constrain the $\omega NN$ coupling, $g_{\omega NN}$, and formfactor cutoff, $\Lambda_{\omega NN}$, we calculate the contribution of $\omega$ $t$-channel exchange to the $\gamma p \to \pi N$ cross section.
We add this contribution to the total proton photoabsorption cross section calculated using the low-density limit of the
$\rho$-meson SF. With a conservative value of $g_{\omega NN} =11$~\cite{Machleidt:1989tm}, we then
find the maximal value of $\Lambda_{\omega NN}$ that renders a total cross section compatible with the data,
see Fig.~\ref{fig:photoabsorption-omega}.
The results for $\Lambda_{\omega NN}=500$ MeV are rather compatible with the data while for the 750 MeV cutoff value
they are sightly higher in the 1100-1300 MeV photon energy range.
Therefore, we choose $\Lambda_{\omega NN}=500$ MeV. For simplicity, we assume this value for all
$\omega N N^*$ formfactor cutoffs.
Figure~\ref{fig:photoabsorption-omega} also shows the individual contribution of the $\omega$ $t$-channel exchange to
the photoabsorption cross section, illustrating its relatively slow growth as a function of photon energy. This contribution
reaches half its maximum value of $\approx11$ $\mu$b at a photon energy of $ E_{\gamma}\approx$ 500 MeV, and its
maximal value at $E_{\gamma} \approx$ 2000 MeV. This suggests that contributions from processes of
$\gamma p \to \pi N^*$ via $\omega$ $t$-channel exchange do not become appreciable until photon energies reach
several hundred MeV higher than the $\pi N^*$ production threshold\footnote{The $\pi \Delta$ production threshold is
irrelevant since the $\omega$ is an isosinglet and cannot excite the nucleon's isospin state.}. The lowest-lying resonance
we consider in this process is the $N(1440)$, which has a $\pi N^*$ production threshold of $\approx 860$ MeV. Therefore,
contributions to resonance production processes via proton photoabsorption that are mediated by an $\omega$ $t$-channel
exchange are negligible in the energy range considered in Fig.~\ref{fig:photoabsorption-omega}.  In principle, we could
also use the process $\gamma p \to \omega p$ to constrain the formfactor cutoff. However, the $\omega p$ production
threshold is $\approx 1100$ MeV, which is too large to be relevant for the photon energy range considered here.

This completes our evaluation of free parameters in our photoemission model. The resulting coupling constants are collected in the Appendix.  We now proceed to calculations of thermal photon emission rates.

\section{Thermal Photon Emission from $\pi\pi$ Cloud}
\label{sec:picloud}
In this section we present our photon emission rates which correspond to modifications of the pion cloud of the
$\rho$ meson. Each process corresponds to an imaginary part of the in-medium $\rho$ selfenergy as shown in
Fig.~\ref{fig:pi-cloud-cuts}. These Born scattering processes are equivalent to the processes contained in the
$\rho$ spectral function. However, instead of using an effective baryon density to approximate the contributions
from all excited baryon states~\cite{Rapp:1999us}, we calculate the contribution from each individual combination
of baryon states explicitly. In order to examine the impact of each process individually, we first display rates for a
temperature of 150 MeV and zero baryon chemical potential, where we multiply the resulting rates by a factor of two
to account for the effect of anti-baryons. In the following, we arrange the discussion of the rates by
partial-wave channel, \ie, whether the $\pi B_1 B_2$ interaction is an $S$-, $P$-, or $D$-wave.

\subsection{$S$-Waves}
\label{ssec:swave}

\begin{figure}
\includegraphics[scale=1.0, center]{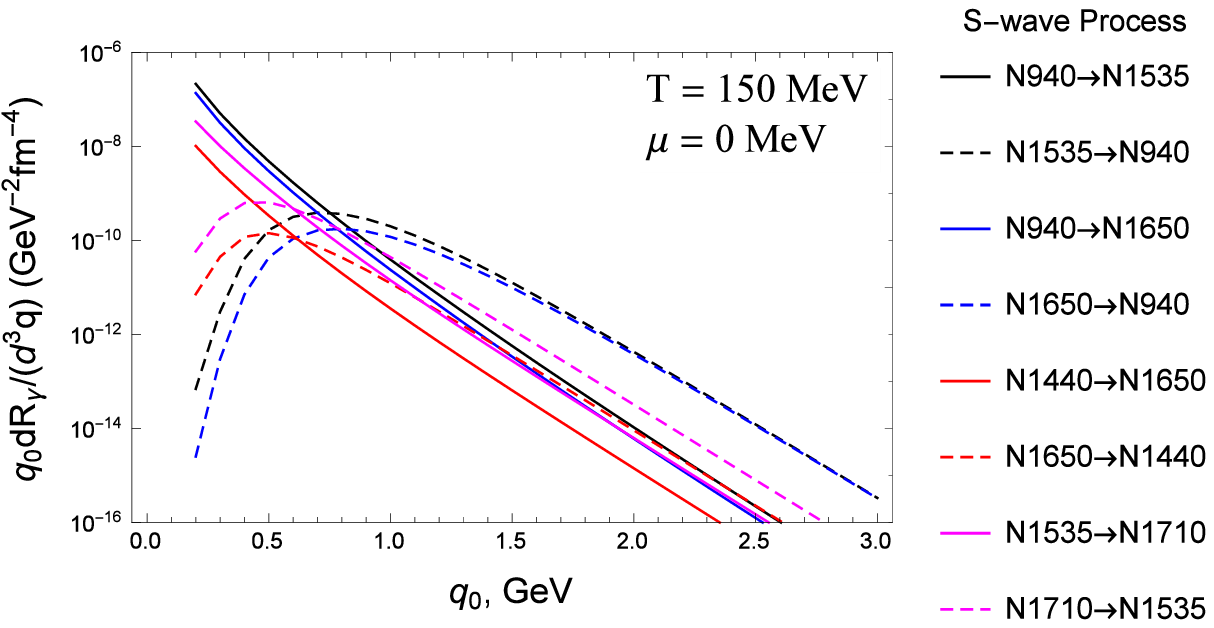}

\vspace{0.5cm}

\includegraphics[scale=1.0,right]{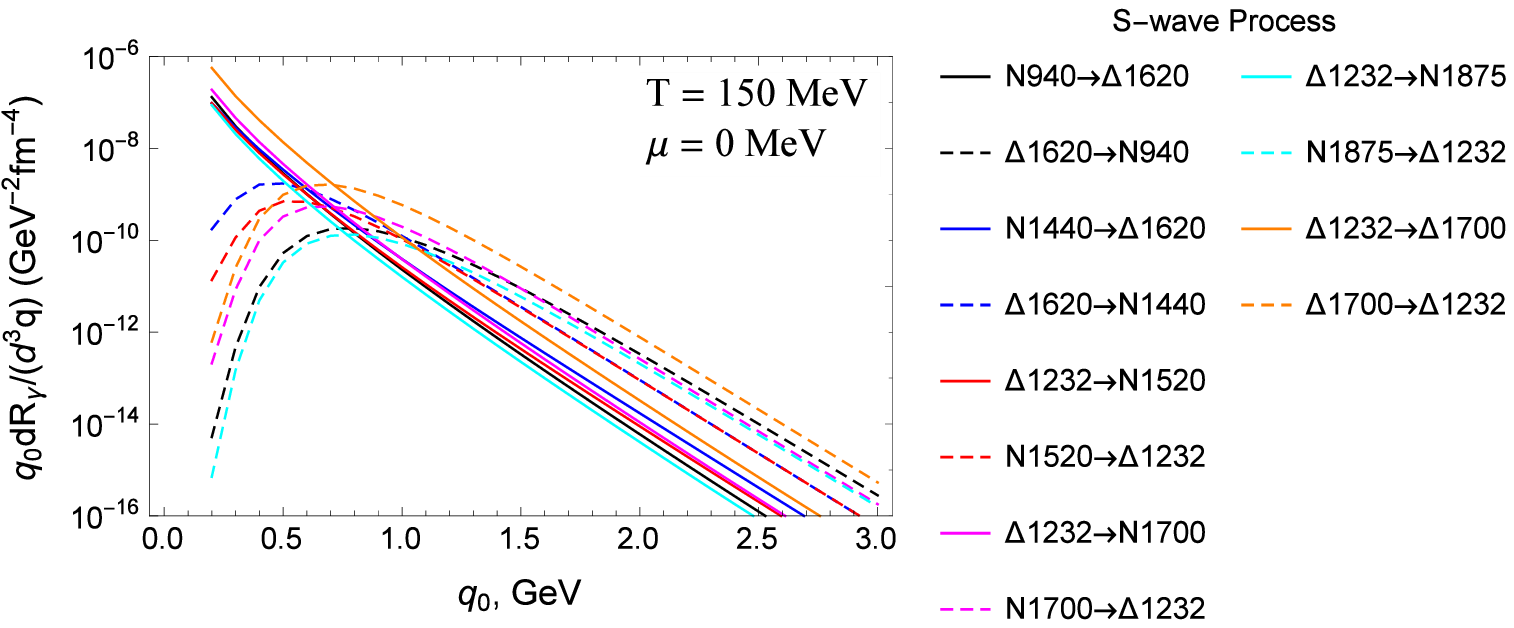}

\vspace{0.5cm}

\includegraphics[scale=1.0, center]{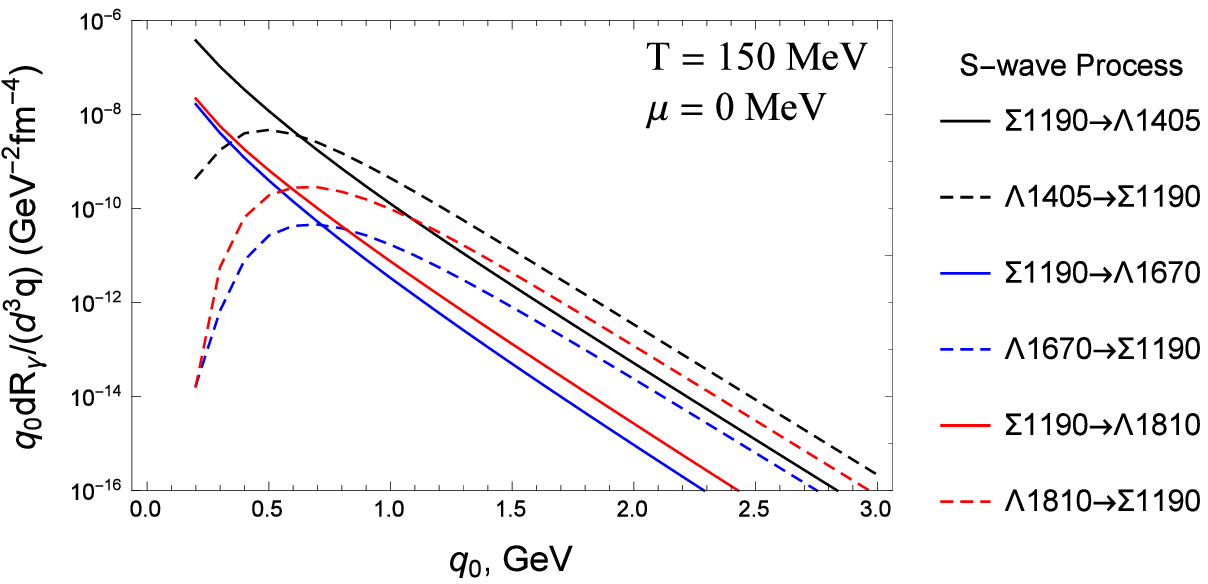}
\caption{Thermal photon production rates from $S$-wave baryon resonances in pion $t$-channel exchange reactions
in $\pi+B_1\to \gamma+B_2$. The upper, middle and lower panel are for processes involving nucleon resonances only,
nucleon and delta resonances, and hyperons, respectively.}
\label{fig:S-wave-results}
\end{figure}

Our $S$-wave results are collected in Fig.~\ref{fig:S-wave-results}.  We immediately see a trend where, in the photon
energy range of $q_0 =0.2-0.5$ GeV, processes involving an initial-state baryon that is less massive than the final-state
one lead to relatively large
photon emission rates. We also see that photon rates from processes involving more massive baryons in the initial state
than in the final state are heavily suppressed in the low-energy range. This is due to the phase space favoring a more
energetic final state when the initial state contains a large amount of extra invariant mass. These processes
dominate at photon energies above $\approx 1$ GeV.

For the processes involving only nucleons and their isospin-1/2 resonances, the low-energy rates are largest for an
incoming nucleon being excited into a $N(1535)$ or $N(1650)$ resonance (both processes also have similar coupling
constants). Basically, the incoming pion can still take advantage
of producing a near on-shell resonance in the final state. On the other hand, for photon energies beyond 1 GeV, the
dominant role is played by processes where those two resonance are incoming with a nucleon in the final state. Here,
the extra invariant mass in the process is effective in producing rather energetic photons. Similar results are seen in
processes involving both nucleon and delta states, in particular the $\Delta(1232)\Delta(1700)$ combination which
dominates at low and high energies, with an incoming and outgoing $\Delta(1232)$, respectively, and to a lesser degree,
the $N(940)\Delta(1620)$ and $\Delta(1232)N(1700)$ combinations.

\subsection{$P$-Waves}
\label{ssec:pwave}
\begin{figure}[H]
\begin{center}
\includegraphics[scale=1.0]{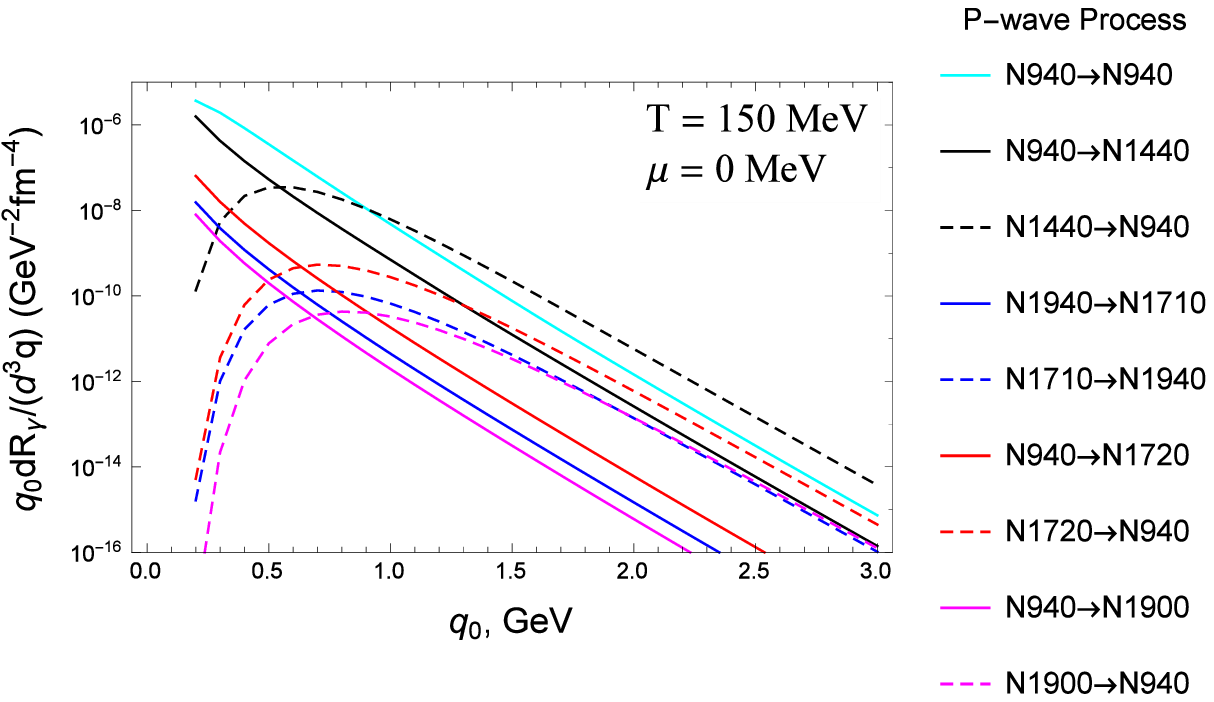}

\vspace{0.5cm}

\includegraphics[scale=1.03]{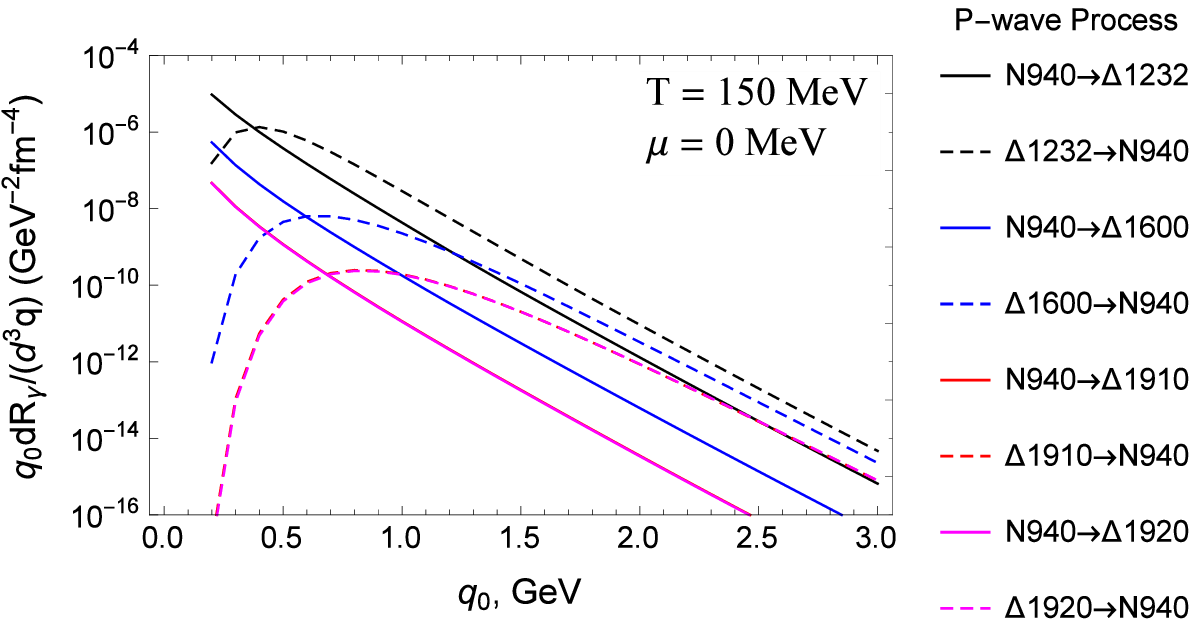}
\end{center}
\caption{Thermal photon production rates from $\pi+B_1\to \gamma+B_2$ involving $P$-wave $\pi N$ couplings
to nucleon resonances (upper panel) and delta resonances (lower panel).}
\label{fig:P-wave-results1}
\end{figure}

\begin{figure}[H]
\begin{center}
\includegraphics[scale=1.0,center]{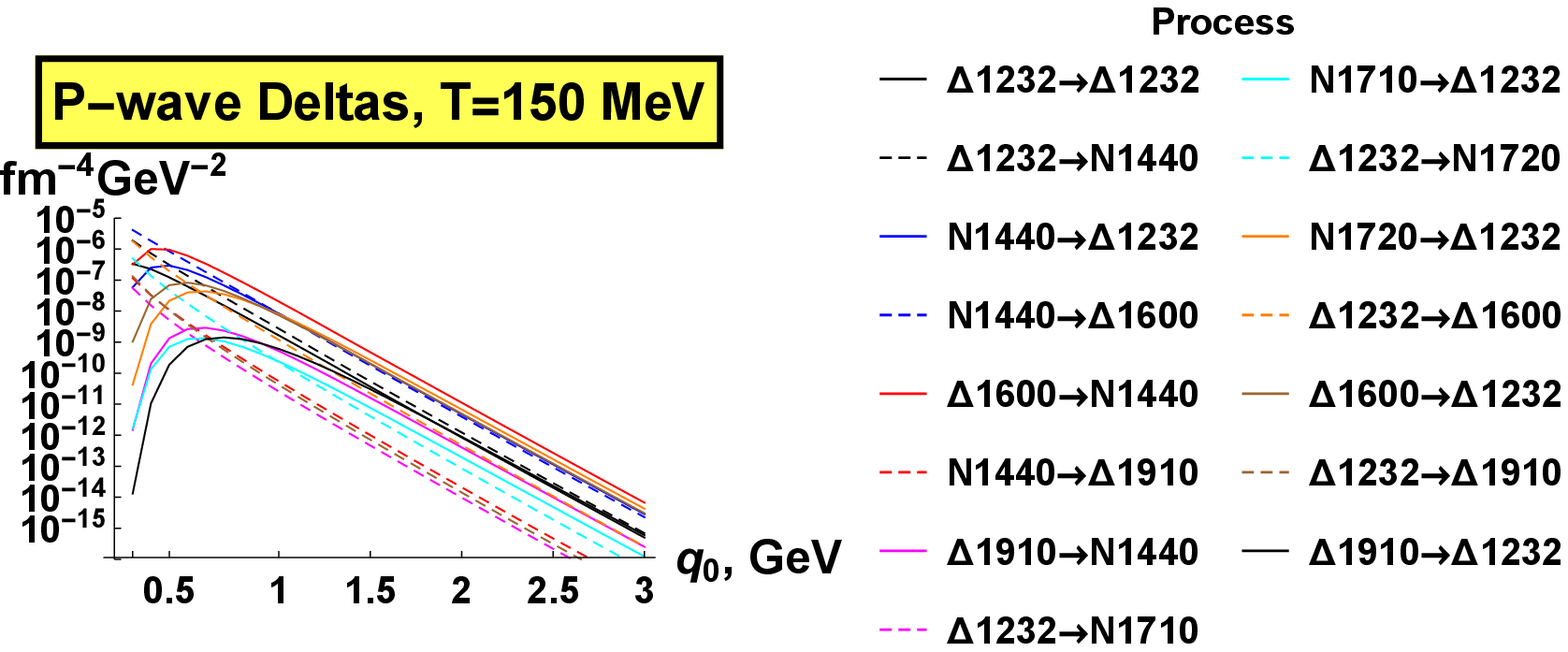}

\vspace{0.5cm}

\includegraphics[scale=1.0]{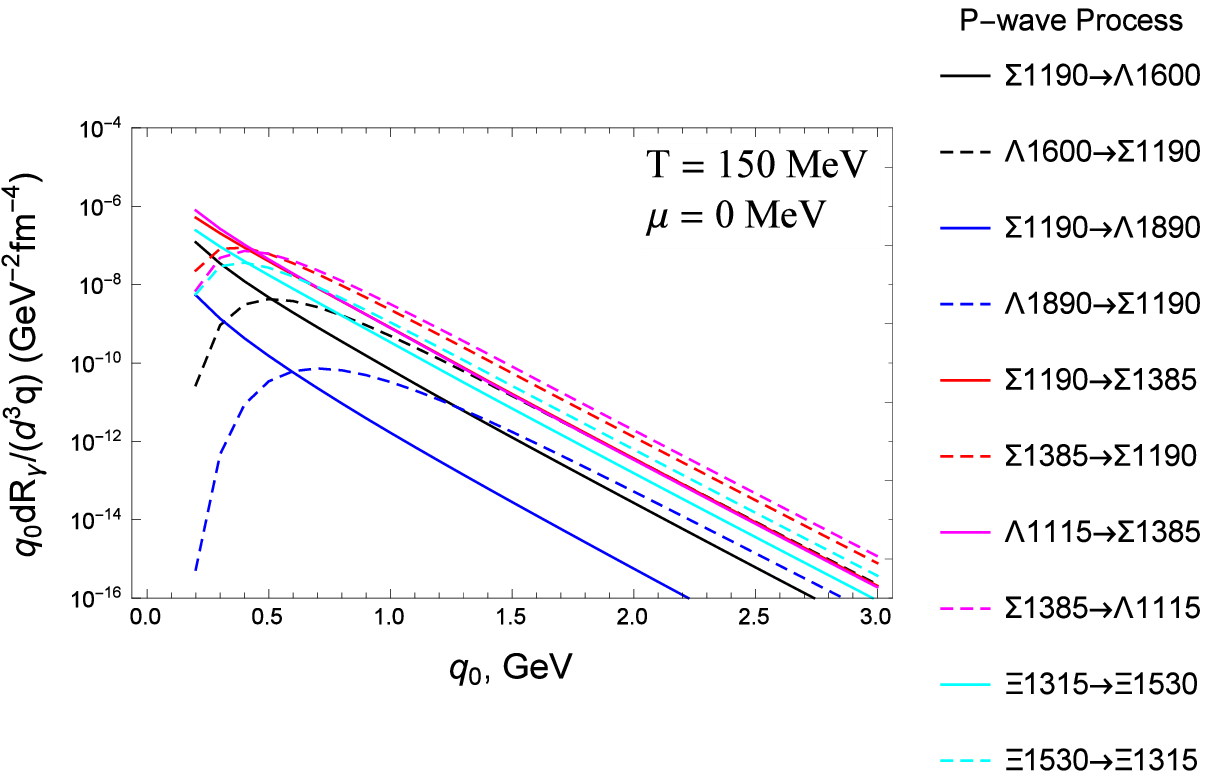}
\end{center}
\caption{Thermal photon production rates from $\pi+B_1\to \gamma+B_2$ involving $P$-wave $\pi B_1B_2$ couplings
for light-flavor resonances without nucleons (upper panel) and for hyperons (lower panel).}
\label{fig:P-wave-results2}
\end{figure}

We now proceed to the $P$-wave processes, displaying the resulting photon rates in Fig.~\ref{fig:P-wave-results1}
(containing all processes involving a ground-state nucleon) and \ref{fig:P-wave-results2} (involving all processes which
do not involve a ground-state nucleon). We first note the greater number of $P$-wave processes as compared to
$S$-wave processes. This alone suggests the $P$-wave interactions will have a greater contribution to the overall rate.
In addition, we also find that the magnitude of the individual $P$-wave rates is generally greater than for $S$-waves.
Inspecting the individual plots, the top panel of Fig.~\ref{fig:P-wave-results1} shows that the $\pi N \to \gamma N$ process is quite
large, as expected (and already included in previous  $\rho$ spectral function calculations). However, the
$\pi N(1440) \to \gamma N$ process (not previously accounted for explicitly) begins to exceed it at
$q_0 \approx 1$ GeV. Two factors contribute to this. First, while the $\pi N N(1440)$ coupling is half that of the $\pi NN$
one, the increased amount of mass in the initial state makes an extra 500 MeV available to be injected into the final
state, although this is somewhat mitigated by the increased suppression from the thermal Fermi factor. Second, the
$\pi NN$ formfactor cutoff is constrained to be 310 MeV, while we recall that the cutoff value for our $\pi B_1 B_2$
formfactors is 920 MeV. This harder formfactor generates less high-energy suppression than in the $\pi N \to \gamma N$
process.

Fig.~\ref{fig:P-wave-results1} displays another expected (and previously calculated) result: processes involving
the $\pi N\Delta$ coupling dominate (with initial nucleon at low $q_0$ and initial delta at high $q_0$). This is mainly
due to the large $\pi N \Delta$ coupling and relatively small masses of the particles, which gives these processes a
generous thermal phase space. Processes with $\pi N\Delta(1600)$ still contribute up to $\approx 50$\% at higher energy,
but only around 10\% at low $q_0$.

The double resonance processes displayed in Fig.~\ref{fig:P-wave-results2} exhibit an unexpected and significant result, namely the magnitude of the rates from $N(1440) \leftrightarrow \Delta(1600)$ processes, which use the
coupling calculated in Sec.~\ref{sssec_cc} using the $N(1440)$ SF. Since the
resulting $\pi N(1440) \Delta(1600)$ coupling is rather large, $f_{\pi N(1440) \Delta(1600)} = 4.9$  (60\% larger than
the $\pi N \Delta$ coupling), the resulting photon emission rates are comparable to the largest ones in the
nucleon sector ($\pi NN$ and $\pi N\Delta$), especially toward higher energies. This result occurs despite the rates being
mitigated by thermal suppression from larger resonance masses. At energies
$q_0\gtrsim 1 $ GeV (\ie, not far from the $\pi N\Delta$ processes) smaller but still significant results are found for processes involving
$\Delta(1232)N(1720)$, $\Delta(1232)N(1440)$, and $\Delta(1232) \Delta(1600)$ exchanges, in good part due to
large spin/isospin factors resulting from spin- and/or isospin-3/2 particles in both the initial and final states.
Finally, we find that the contributions from $P$-wave hyperons are overall
smaller than in the light-flavor sector, by almost an order of magnitude, but are not entirely negligible.

\subsection{$D$-Waves}
\label{ssec:dwave}
\begin{figure}[H]
\begin{center}
\includegraphics[scale=1.0]{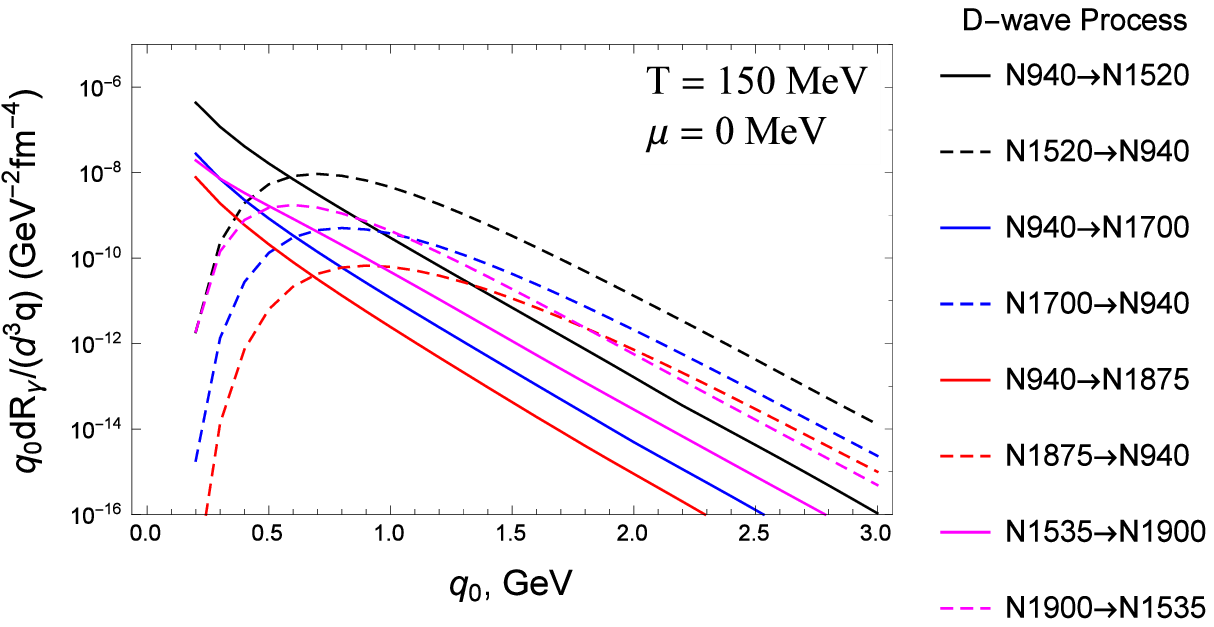}

\includegraphics[scale=1.0]{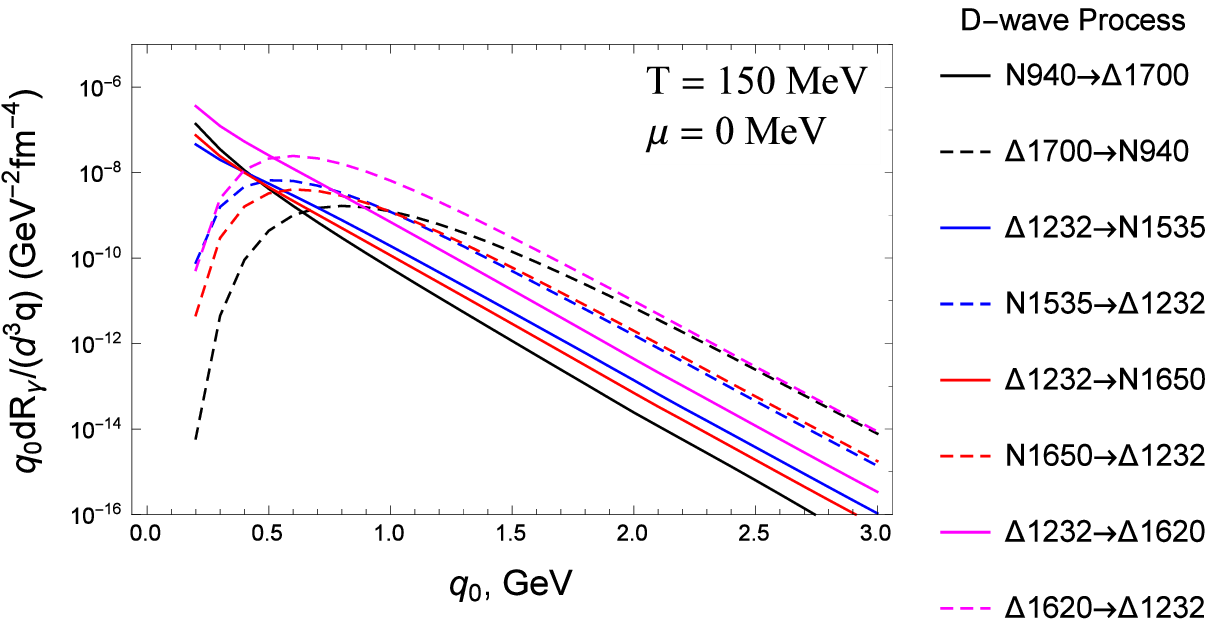}

\includegraphics[scale=1.0]{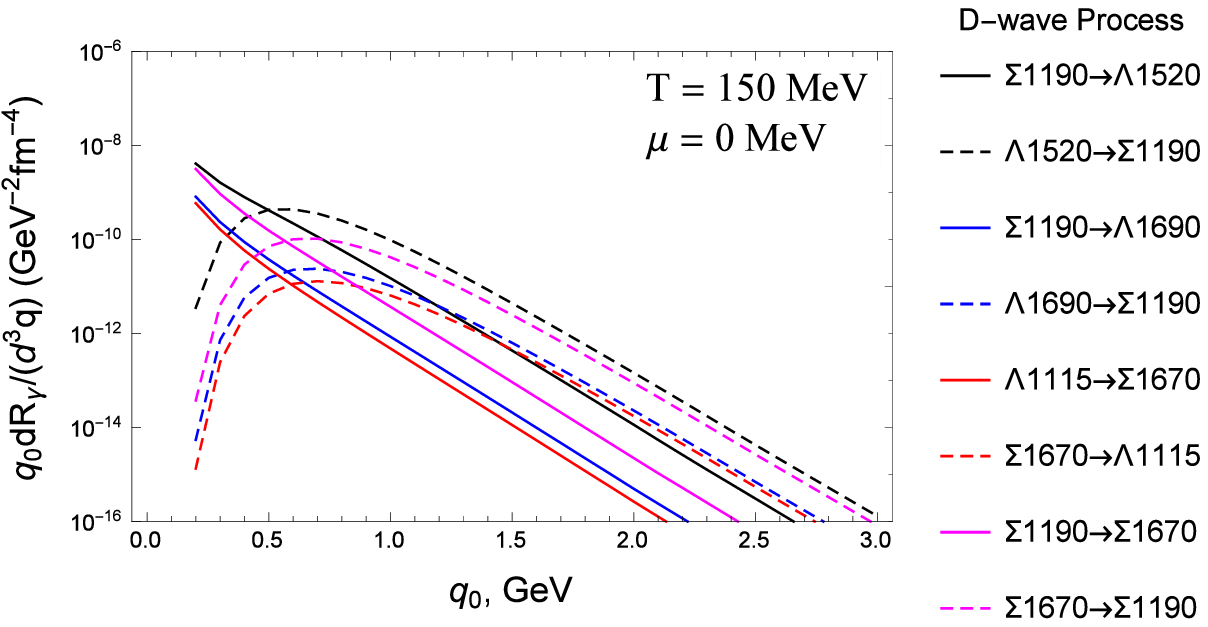}
\end{center}
\caption{Thermal photon production rates from $\pi+B_1\to \gamma+B_2$ involving $D$-wave $\pi B_1B_2$ couplings
for light-flavor resonances (upper panels) and hyperons (lower panel).}
\label{fig:D-wave-results}
\end{figure}

Our final contribution from the $\rho$-meson's $\pi \pi$ cloud is due to processes from $D$-wave pion-baryon couplings.
The corresponding rates are summarized in Fig.~\ref{fig:D-wave-results}.
In the light-flavor sector of nucleon and delta states the largest rates are those involving $N(940)N(1520)$ and
$\Delta(1232)\Delta(1620)$ exchanges.  At the highest photon energies of near 3\,GeV considered here they are very
comparable to (and even slighter larger than) the largest $P$-wave sources discussed previously. However, in the
phenomenologically more relevant region of $q_0\simeq1-2$\,GeV, they are somewhat smaller than the leading $P$-wave
induced emission rates. Additionally, the number of individual channels in the latter largely outnumber those in the former, thereby reducing the overall impact of $D$-wave processes.
The harder spectra of the $D$-wave rates can be attributed to the higher power of momentum
in the pertinent vertices as compared to the $P$-wave (and even more so $S$-wave). This implies that phase space
effects lead to substantial deviations of the rates' spectral slopes from a purely thermal behavior. Other
``distortions'' away from the thermal behavior are the effects of formfactors and the different combinations of baryon
masses in the entrance and exit channels as discussed above.
As for the processes with $D$-wave hyperons, their rates are much smaller than those from the nucleons or deltas by
typically two orders of magnitude, and are thereby largely negligible.

\subsection{Total $\pi\pi$ Cloud Rates}
\label{ssec:pipi-total}
\begin{figure}[H]
\begin{center}
\subfloat{
\includegraphics[scale=1.3]{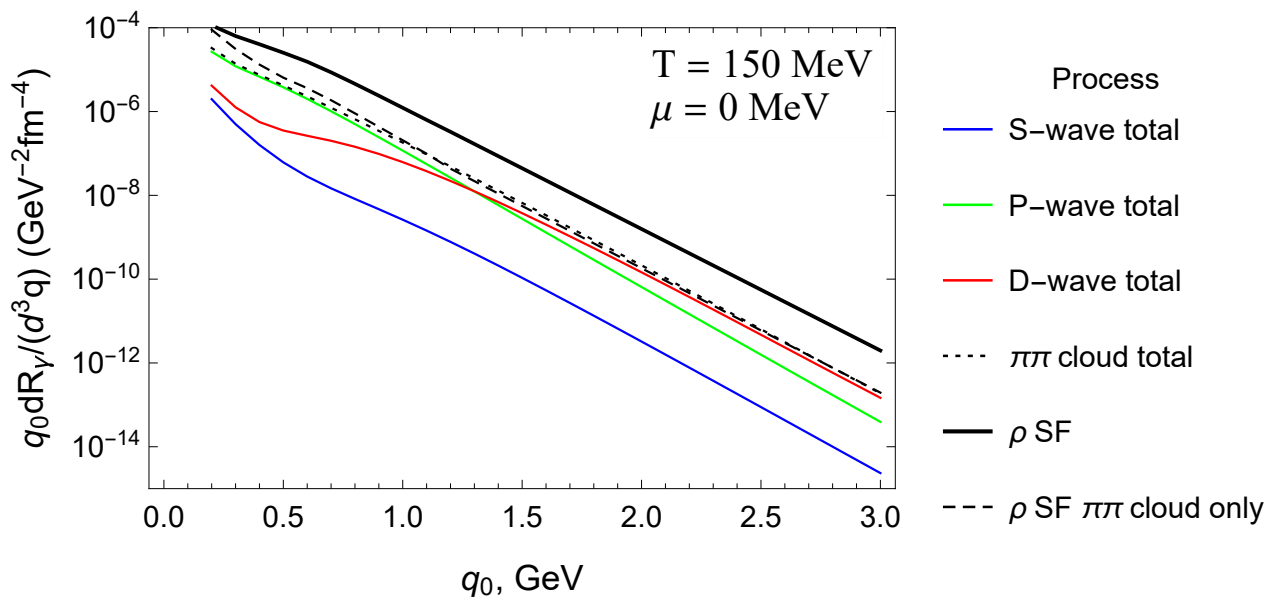}}
\end{center}
\caption{Thermal photon rates from $\pi+B_1\to \gamma+B_2$ reactions with pion exchange calculated in this work in the
$S$-wave (blue line), $P$-wave (green line) and $D$-wave (red line) channels,  and their sum (black dotted line),
compared to the pion cloud contribution from the $\rho$ SF~\cite{Rapp:1999us} including only $N$ and
delta states but using an effective nucleon density (black dashed line), and the total rate from the $\rho$ spectral
function (black solid line). }
\label{fig:pioncloud-results}
\end{figure}

The total pion cloud contributions by partial wave, as well as their sum, are compiled in Fig.~\ref{fig:pioncloud-results}.
We compare these rates to those from the in-medium $\rho$ spectral
function~\cite{Urban:1998eg,Urban:1999im,Rapp:1999us}, both for the total and for the pion cloud component only,
evaluated with an effective total nucleon density of 0.13$\rho_0$ at $T=150$ MeV and $\mu_B=0$~\cite{Rapp:2000pe}
to approximately account for higher resonances and anti-baryons.
Our explicit calculation involving 25 baryon resonances results in a total rate which is quite close to the one from the
$\rho$ SF with only the ground-state nucleon and delta states but with an upscaled effective nucleon
density estimated as the nucleon density plus half of the density of all excited states (including deltas and hyperons)
in Ref.~\cite{Rapp:1999us}. There is some discrepancy at low photon energies, below about 1\,GeV, where the
SF-based results are larger, possibly due to resummation effects or the finite width included in the in-medium delta
and nucleon propagators.
However, at vanishing baryon chemical and $T=150$ MeV potential the pion cloud contribution
only makes up $\approx 17\%$ of the total rates from the $\rho$ SF at $q_0=1.0$ GeV.
This implies that under these conditions the photon rates from are dominated by mechanisms other than the $\pi\pi$ cloud,
mostly radiative decays of meson (and to a lesser extent baryon) resonances, as well as other mesonic reactions which
are not part of the SF, including $\omega$ $t$-channel exchange in $\pi\rho\to\pi\gamma$. The importance
of the latter channel motivates us to scrutinize similar reaction in the baryonic sector, which will be done in the next section.

\section{Photon Emission from $\pi \omega$ Cloud}
\label{sec:pi-om-cloud}
In this section we investigate the contributions from modifications to the $\pi \omega$ cloud of the $\rho$ meson. Other than
$\omega N \to \gamma N$, these processes are thus far unexplored for thermal photon rates; specifically, they are not
contained in in-medium $\rho$ selfenergy calculations. As mentioned above, due the structure of the $\pi \rho \omega$
vertex, additional diagrams are not required to ensure gauge invariance. This simplifies the
photoemission calculations as we need only calculate the dominant $t$-channel $\omega$ exchange processes.

\subsection{$\omega$ Exchange Processes}
\label{ssec:omexch}
\begin{figure}[H]
\begin{center}
\includegraphics[scale=1.0]{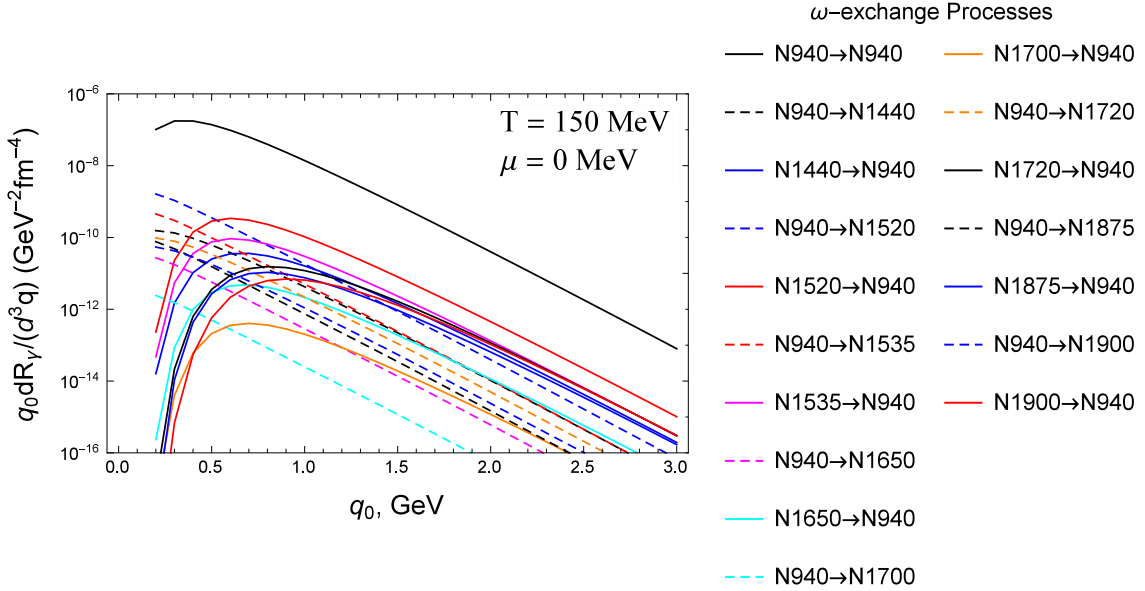}
\end{center}
\caption{Thermal photon emission rates from $\pi B_1 \to \gamma B_2$  processes involving $t$-channel $\omega$ exchange.}
\label{fig:om-rate-1}
\end{figure}

Photon production from pion-baryon scattering with an $\omega$ $t$-channel exchange corresponds to cuts of the
$\pi \omega$ cloud of the $\rho$ SF as shown in Fig.~\ref{fig:omega-cloud-cuts}(a).  Since the emitted
photon is attached to the $\pi \rho \omega$ vertex, these processes, which involve the $\omega N N^*$ vertices are
stand-alone gauge invariant, without the need to consider further diagrams. In the present work we could only
estimate the coupling constants for nine $N^*$ resonances, so the baryon spectrum is not as extensive as for $\pi N N^*$
couplings.
The resulting rates for the $\omega$ $t$-channel exchanges are shown in Fig.~\ref{fig:om-rate-1}. The process with
two nucleons produces a rate which is rather comparable to that from leading pion-exchange channels, but the
magnitudes of the $\omega N N^*$ coupling constants (which are all less than 3) are much reduced relative to
$g_{\omega NN}=11$ which dwarfs all other processes (the rates depend quadratically on the coupling). Since the
rates from the resonance processes with $\omega$ exchange are all three or more orders of magnitude smaller than
the total pion exchange contribution, we will neglect them and keep only the contribution from the $\pi N \to \gamma N$
$\omega$ $t$-channel process in our final results.

\subsection{External $\omega$ Mesons}
\label{ssec:omext}

\begin{figure}[H]
\centering
\includegraphics[scale=0.4]{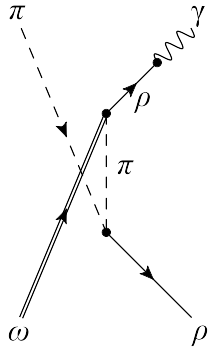}

\vspace{0.5cm}

\caption{Feynman diagram of photon emission from $u$-channel pion exchange in the $\pi\omega\to\gamma\rho$ scattering
process as considered in Ref.~\cite{Holt:2015cda}.}
\label{fig:piomega}
\end{figure}

The second contribution from  the cut of the $\pi \omega$ cloud of the $\rho$ SF, as shown in
 Fig.~\ref{fig:omega-cloud-cuts}(b),  corresponds to scattering processes of the type
$\omega B_1 \to \gamma B_2$ via $t$-channel pion exchange. As with the diagrams involving a $t$-channel
$\omega$ exchange, these processes involve photon emission from the $\pi \rho \omega$ vertex, and are thus
gauge invariant by themselves. They are topologically similar to the processes generated by the $\pi\pi$ cloud and only
involve swapping the $\rho \pi \pi$ vertex for the $\pi \rho \omega$ vertex. This yields a new set of
processes, all involving the same combinations of baryons as considered in Sec.~\ref{sec:picloud}. While these
new processes are suppressed by the $\omega$ mass in the exchange propagator, they also receive a significant
boost from the large size of the $\pi \rho \omega$ coupling constant. We therefore anticipate their contribution
to the overall photon rate to be appreciable.

We note that these processes have the same topological configuration as the $u$-channel pion exchange diagram in the
$\pi \omega \to \gamma \rho $ process (shown in Fig.~\ref{fig:piomega}), whose contribution to thermal photon production
was calculated in Ref.~\cite{Holt:2015cda}. In that work it was found that is was kinematically allowed for the exchanged
pion to go on-shell. This resulted in both a non-integrable singularity in the photon emission integral and a double-counting
of the $\omega$ radiative decay, which was already included in the in-medium $\rho$ SF of
Refs.~\cite{Urban:1999im,Rapp:1999us}. The same on-shell behavior shows itself in Fig.~\ref{fig:omega-cloud-cuts}(b).
If the incoming baryon is less massive than the outgoing baryon by at least the pion mass, the exchanged pion can go on
shell. We then have two separate processes for the $\omega \to \pi^0 \gamma$ radiative decay and a $B_1 \to \pi B_2$
absorption, all involving on-shell particles. However, using a correspondence between thermal field theory and relativistic
kinetic theory, it was found in Ref.~\cite{Holt:2015cda} that the problem can be dealt with in  thermal field theory by
excluding the Landau cut of the $\rho / \gamma$ selfenergy. In the kinetic theory formalism this is equivalent to
restricting the energy range of the incoming $\omega$ particle such that it is less energetic than the outgoing photon.
We use that procedure here by applying the same kinematic restriction of $E_{\omega} < q_0$ to the integration range
of the photon emission integral. This precludes the possibility of the exchanged pion going on-shell.

\begin{figure}[H]
\centering
\includegraphics[scale=1.3]{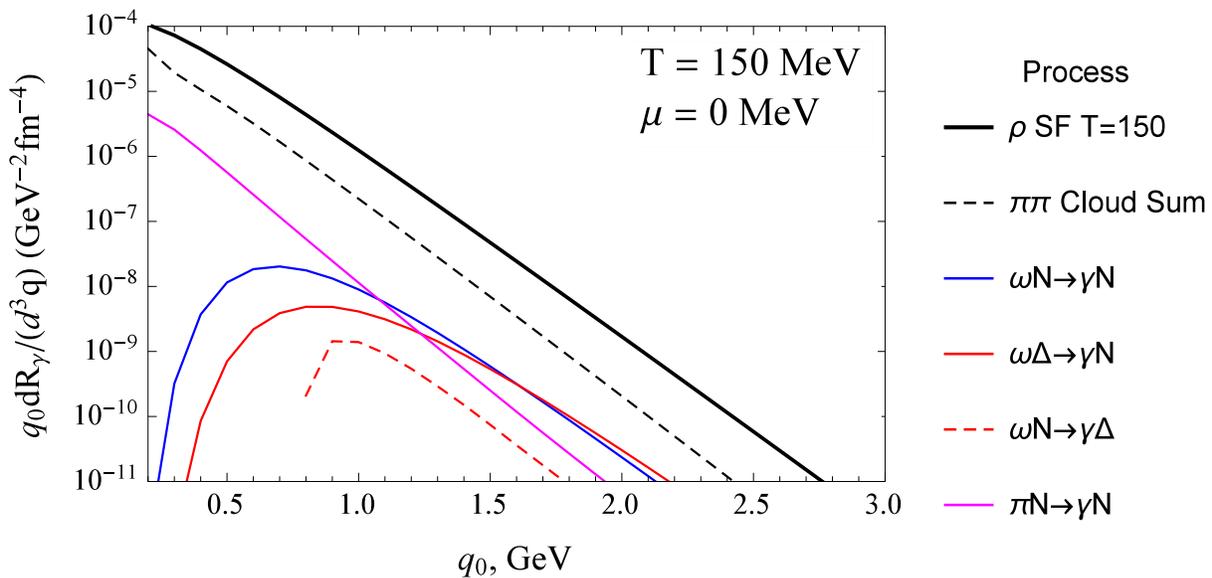}
\caption{Thermal photon rates at $T=150$ MeV and $\mu_B=0$ from processes with an external $\omega$
particle scattering off baryons (red and blue solid lines, as well as red dashed line), compared to the $\pi N\to\gamma N$
process with pion exchange (solid pink line), the total pion cloud rate (black dashed line) and the total rate from the
medium $\rho$ SF from Ref.~\cite{Rapp:1999us} .}
\label{fig:omega-ext-rate-sample}
\end{figure}

Since we again have a considerable number of processes, we will only plot the rates of the ones with the lightest external
baryons, then analyze the total. The former are displayed in Fig.~\ref{fig:omega-ext-rate-sample}. We first note the effect
of the kinematic restriction, $E_{\omega} < q_0$, which removes the low-$q_0$ range from the $\omega N \to \gamma \Delta$
process, causing it to have no contribution for photon energies less than 1 GeV. This is the same behavior displayed by the
kinematic restriction (or equivalently, the Landau cut) in Fig.~4 of Ref.~\cite{Holt:2015cda}.  Second, we note the sizes of
the individual processes. For comparison, we plotted rates from $\pi N \to \gamma N$ scattering as calculated in
Sec.~\ref{sec:picloud}.  In the phenomenologically relevant range around $q_0 \approx 1$ GeV, the $\omega N \to \gamma N$
rate is comparable to the $\pi N \to \gamma N$ one, suggesting its possible relevance in the overall rates. This is due to
the large size of the $\pi \rho \omega$ coupling constant overcoming the increased thermal suppression of the
$\omega$ as an external particle as compared to the pion. Third, we see that at energies above $q_0 \approx 1$ GeV,
the incoming $\omega$ processes rapidly gain strength. This indicates that the total rate from the external $\omega$
processes is rather significant at high photon energies.

\section{Total Rates and Comparison to Existing Calculations}
\label{sec:results}
\begin{figure}[H]
\begin{center}
\includegraphics[scale=1.2]{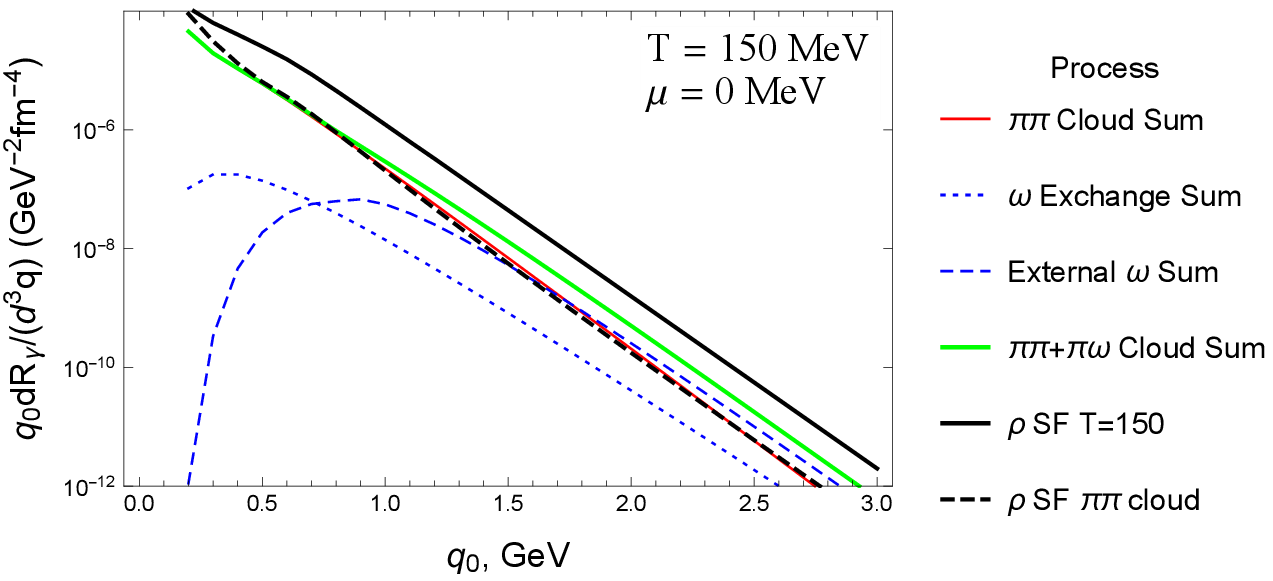}

\vspace{0.5cm}

\includegraphics[scale=0.85]{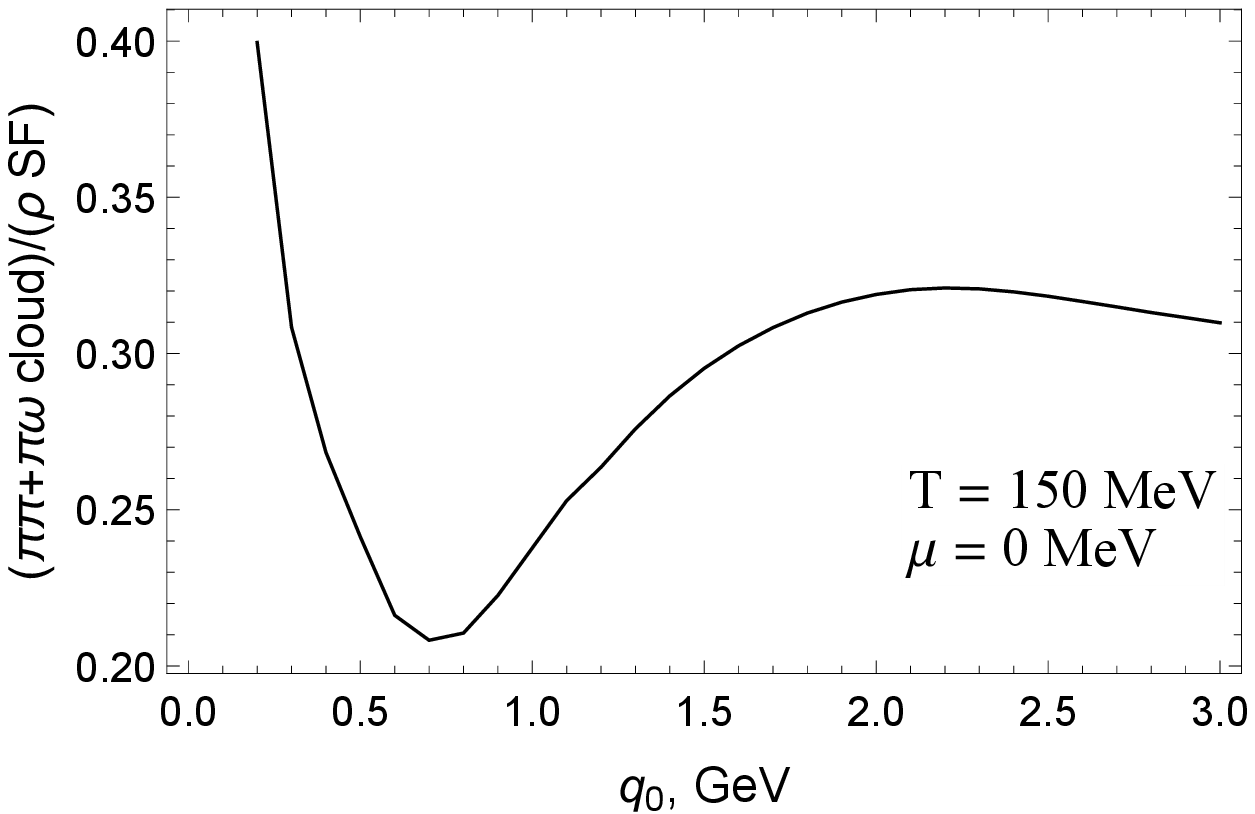}
\end{center}

\vspace{0.5cm}

\caption{Upper panel: Total thermal photon rates at $T=150$ MeV and vanishing baryon chemical potential calculated in this
work from the $\pi \pi$ cloud (red line) and the $\pi\omega$ cloud (blue dotted and dash-dotted lines); their sum (green solid
line) is compared to the total rate from the in-medium $\rho$ SF (solid black line) as well as its $\pi\pi$ cloud component only (black dashed line). Lower panel: Ratio of rates from
$\pi\pi$ plus $\pi\omega$ cloud as calculated in this work (corresponding to the green line in the upper panel) to the total
rates from the in-medium $\rho$ SF (black solid line in the upper panel). }
\label{fig:totals}
\end{figure}

The final step is to evaluate the relevance of our total rates photon from the $\pi\omega$  and $\pi\pi$ cloud in
comparison to existing results, specifically, to the total rate from the in-medium $\rho$ SF of
 Ref.~\cite{Rapp:1999us}. We first focus on conditions of a net-baryon free system, for a temperature of
$T=150$ MeV and $\mu_B=0$, \textit{cf}.~Fig.~\ref{fig:totals}.
As anticipated earlier, the effect of the $\pi\omega$ cloud
processes is quite significant at photon energies $q_0 \gtrsim 1.0$ GeV (\textit{cf.} the green solid line in the upper panel
of Fig.~\ref{fig:totals}). For example, at $q_0=1$ GeV, the inclusion of
$\pi\omega$ cloud effects lifts the total from 17\% with just the $\pi\pi$ cloud to 24\%, while  at $q_0=2$ GeV
the increase is from 12\% to 32\%, and at $q_0=3$ GeV it is from 8\% to 31\% (\textit{cf}.~the lower panel of
Fig.~\ref{fig:totals}). Therefore, for photon energies
above 1 GeV, the effect of the $\pi \omega$ cloud is substantial.

Next we inspect the size of the contribution of our $\pi\pi$ cloud calculations relative those of the $\rho$ spectral
function in a more baryon-rich environment, by increasing the baryon chemical potential to  $\mu_B =340$\,MeV
at the same temperature of  $T=150$\,MeV. In Ref.~\cite{Turbide:2003si}, the left-hand panel
of  Fig.~3 separates out the individual contributions to the SF under these conditions.
At a photon energy of $q_0 =1$\,GeV,
we take the difference of the value of the full spectral function and the SF with no baryons.
This gives us the baryonic contribution whose value is approximately $1\times 10^{-6}$\,fm$^{-4}$\,GeV$^{-2}$.
This is split approximately evenly between pion cloud effects and direct $\rho BB$ interaction effects. Therefore, the
estimated contribution from the pion cloud is $5\times10^{-7}$\,GeV$^{-2}$. A direct calculation of our pion cloud rates
at $T=150$\,MeV and $\mu_B =340$\,MeV at the same photon energy yields a rate of $9.76\times10^{-7}$\,GeV$^{-2}$.
This indicates that at $q_0=1$\,GeV we have found a 100\% enhancement of pion cloud rates, which is a 50\%
enhancement of baryonic rates, resulting in a net 25\% enhancement of the photon rates given by the in-medium
$\rho$ SF. We note however, that in the latter the pertinent selfenergies are added
coherently, while our kinetic-theory results correspond to an incoherent sum of all processes. The extent to which this difference
depends on this effect warrants further scrutiny.

\begin{figure}[H]
\begin{center}
\includegraphics[scale=1.2]{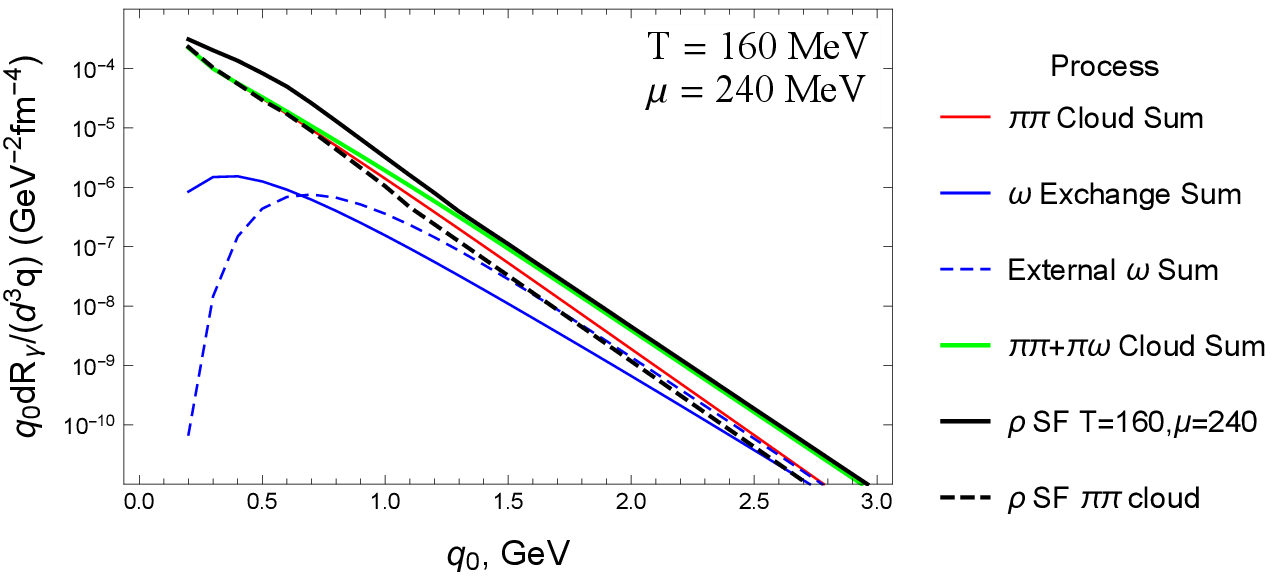}

\vspace{0.5cm}

\includegraphics[scale=0.85]{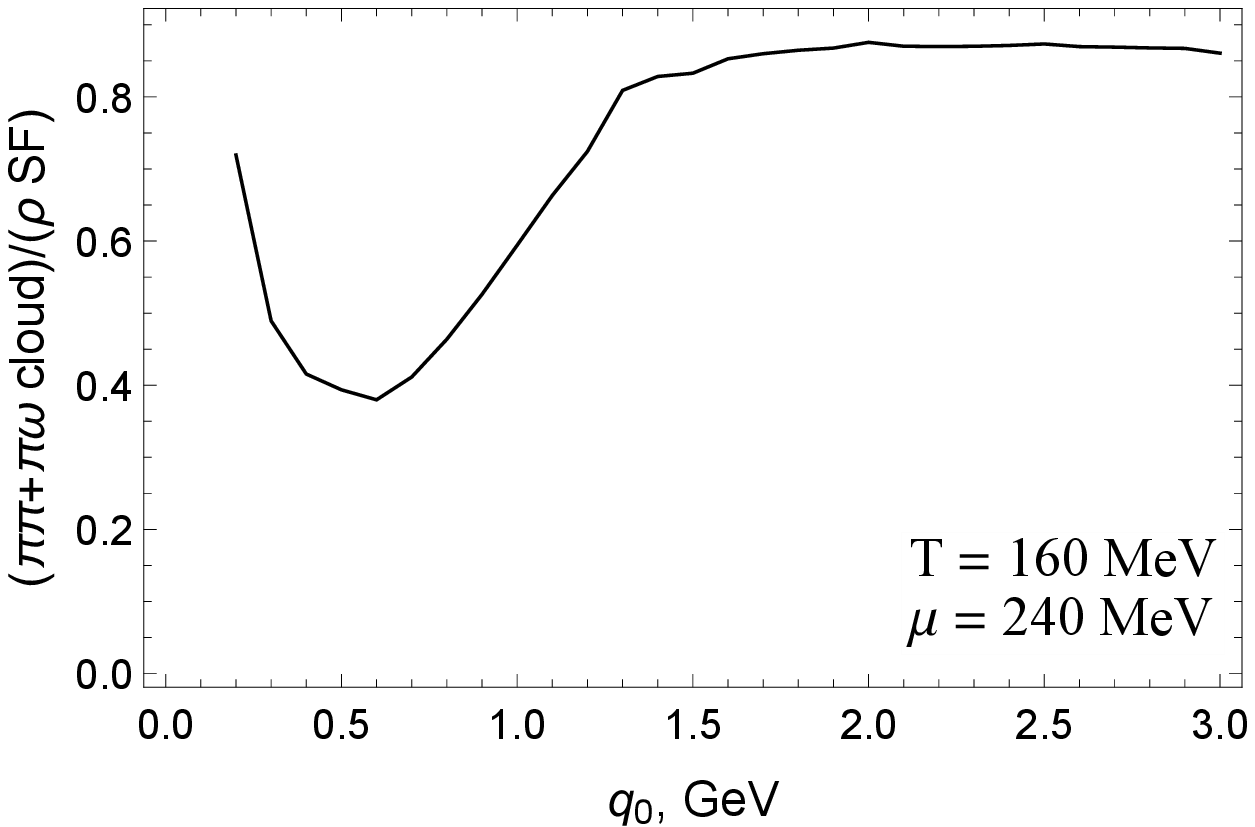}
\end{center}

\vspace{0.5cm}

\caption{Upper panel: Total thermal photon rates at typical chemical freezeout conditions of $T=160$ MeV and $\mu_B=240$ MeV
from our calculations of the $\pi \pi$ cloud (red line), $\pi\omega$ cloud (blue solid and dashed lines); their sum
(green line) is compared to the total rate from the in-medium $\rho$ SF (solid black line) as well as its $\pi\pi$ cloud component only (black dashed line).
 Lower panel: Ratio of rates from
$\pi\pi$ plus $\pi\omega$ cloud as calculated in this work (corresponding to the green line in the upper panel) to the total
rates from the in-medium $\rho$ SF (black solid line in the upper panel).}
\label{fig:totals-160240}
\end{figure}

Finally, let us compare our results to the rates from the $\rho$ SF at chemical-freezeout
conditions at the top center-of-mass energy available in Pb-Pb collisions at the Super Proton Synchrotron (SPS,
($\sqrt{s}=17.3$\,GeV), $T=160$ MeV and $\mu_B=240$ MeV, \textit{cf}.~Fig.~\ref{fig:totals-160240}. Here,
the total baryon density amounts to 0.776\,$\rho_0$ with a nucleon density of 0.204\,$\rho_0$. Following
the prescription of an effective nucleon density of Ref.~\cite{Rapp:1999us} one obtains
$\rho^{\rm eff}_N = \rho_N + \rho_{B*} = 0.51\rho_0$, which also contains a 5\% increase from anti-baryons.
This is approximately a factor of four larger than the value of 0.13$\rho_0$ at $\mu_B=0$ and $T=150$\,MeV.
Consequently, relative to  Fig.~\ref{fig:totals}, the effects of baryons in the $\rho$ SF
increases substantially. We first note that our calculated $\pi\pi$ cloud contribution (red solid) line now shows
good agreement with the SF calculations at low photon energies, while it becomes slightly larger for energies above
$\approx 0.8$ GeV. In addition, the process $\pi N \to \gamma N$ with
$\omega$ $t$-channel exchange has markedly increased relative to the total rates, so that our total calculated
$\pi\pi$ plus $\pi\omega$ cloud effects are now quite comparable to the total $\rho$ SF contribution for energies
above $\approx 1.3$ GeV This suggests that our newly calculated  contributions to thermal photon emission further
increase the significance of baryonic sources at higher densities.

\section{Conclusions}
\label{sec:concl}
In this work we have calculated thermal-photon emission rates from hadronic matter through  $2\to2$ pion-baryon and
$\omega$-baryon Born scattering processes using a rather extensive set of 15 light-flavor resonances (plus nucleons)
and 11 hyperons, including various mutual couplings. We have considered most of the reaction channels for which the
particle data group lists significant coupling strength for $\pi$-baryon-baryon and $\omega$-baryon-baryon vertices,
leading to a total of about 60 channels. The pertinent coupling constants have been constrained by
the relevant baryon decay branchings, and combined with additional information on the momentum dependence through
vertex formfactors constrained by pion-nucleon scattering data and photoabsorption cross sections on the nucleon.
Electromagnetic gauge invariance has been enforced through the Pauli-Villars scheme in the pion sector while processes
involving the $\omega\rho\pi$ vertex are either gauge-invariant by themselves or have been treated by a previously used
formfactor averaging procedure. While our rate calculations have been carried out within the kinetic-theory framework, we
have emphasized their intimate connection to thermal field-theory calculations of the $\rho$-meson spectral function.
In the latter framework, the processes that we have calculated correspond to medium modifications of the $\pi\pi$ and
$\pi\omega$ cloud of the $\rho$ meson.

For the pion-baryon processes, we have found that the dominant contributions to photon emission are generated by
$P$-wave processes up to energies of about 1\,GeV. At higher energies $D$-wave processes become very
comparable, even slightly larger in the rate, while the $S$-wave contributions are more than an order of magnitude smaller.
Among the individual processes, the importance of the previously considered delta- and nucleon-induced processes
was confirmed, but notable new contributions of comparable magnitude were identified as being due to $NN(1440)$
couplings, $\Delta(1232)$ couplings to $N(1440)$, $\Delta(1600)$, and $N(1720)$, as well as the $N(1440)\Delta(1600)$
channels.
When comparing our total results for pion exchange calculations to existing $\rho$ spectral function calculations using only
$N\Delta$ states but with a previously defined effective nucleon density, we found fair agreement. Some discrepancies
at low photon energies were found, which might be due to resummation effects in the $\rho$ selfenergy which warrant
further investigation. Our microscopic calculations carried out here largely validate the approximate rates used in numerous
phenomenological applications to date and put our understanding of the role of excited baryons for the EM emissivity of
hot and dense hadronic matter on a firmer basis.

The second part of our study was devoted to processes with $\omega$ meson-baryon couplings, most of which constitute
novel sources of thermal photon production. These turn out to be significant relative to the pion cloud contributions for
photon energies of around 1 GeV (several tens of percent), and exceed those for energies above $\approx 1.5$ GeV.
The $\omega$-induced rates also have potential for additional contributions, as many of their manifestations in baryon decay
branchings are likely not well established or not even known at all yet.

Directions for future work include applications of our results in calculations of thermal-photon spectra in
heavy-ion collisions, where they could help to mitigate current tensions with experimental data. The implementation
of our Born diagrams into the $\pi\pi$ and a newly introduced $\pi\omega$ cloud of the in-medium $\rho$ meson
are also of considerable interest, not only toward a more complete description of EM emission but to study interference
effects in the (resummed) propagators in the $\rho$ selfenergy that we have not assessed within the kinetic-theory
framework employed in the present paper.

\appendix*
\section{Particle Information}
\label{sec:app}

\begin{table}
\begin{center}
\begin{tabular}{|c|c|c|c|c|c|}
\hline
Parent & \begin{tabular}{@{}c@{}}Full \\ Width (MeV) \end{tabular} & \begin{tabular}{@{}c@{}}Decay \\ Mode \end{tabular} & \begin{tabular}{@{}c@{}}Partial \\ Width (\%) \end{tabular} & \begin{tabular}{@{}c@{}}Partial \\ Wave \end{tabular} & \begin{tabular}{@{}c@{}} $p_{CM}$ \\ (MeV) \end{tabular} \\
\hline
$N(1440)$ & 350 & $N\pi$ & 65 & P & 391 \\
\hline
  & 350 & $\Delta \pi$ & 20 & P & 135 \\
\hline
$N(1520)$  & 115 & $N\pi$ & 60 & D & 453 \\
\hline
   & 115 & $\Delta\pi$ & 15 & S & 225 \\
\hline
$N(1535)$   & 150 & $\Delta\pi$ & 2 & D & 242 \\
\hline
    & 150 & $N\pi$ & 45 & S & 468 \\
\hline
$N(1650)$   & 140 & $\Delta\pi$ & 12.5 & D & 344 \\
\hline
    & 140 & $N\pi$ & 60 & S & 546 \\
\hline
   & 140 & $N(1440)\pi$ & 3 & S & 147 \\
\hline
$N(1700)$   & 150 & $\Delta\pi$ & 50 & S & 385 \\
\hline
   & 150 & $N\pi$ & 12 & D & 580 \\
\hline
$N(1710)$   & 125 & $N\pi$ & 12.5 & P & 587 \\
\hline
    & 125 & $\Delta\pi$ & 39 & P & 393 \\
\hline
    & 125 & $N(1535)\pi$ & 15 & S & 100 \\
\hline
$N(1720)$   & 250 & $N\pi$ & 11 & P & 593 \\
\hline
   & 250 & $\Delta\pi$ & 75 & P & 401 \\
\hline
$N(1875)$   & 250 & $N\pi$ & 7 & D & 694 \\
\hline
    & 250 & $\Delta\pi$ & 40 & S & 520 \\
\hline
$N(1900)$   & 200 & $N\pi$ & 5 & P & 710 \\
\hline
   & 250 & $\Delta\pi$ & 7 & D & 305 \\
\hline\hline
$\Delta$ & 117 & $N\pi$ & 100 & P & 229 \\
\hline
$\Delta$(1600) & 320 & $N\pi$ & 17.5 & P & 513 \\
\hline
    & 320 & $\Delta\pi$ & 55 & P & 328 \\
\hline
    & 320 & $N(1440)\pi$ & 22.5 & S & 75 \\
\hline
$\Delta$(1620)    & 140 & $N\pi$ & 25 & S & 534 \\
\hline
    & 140 & $\Delta\pi$ & 45 & D & 318 \\
\hline
    & 140 & $N(1440)\pi$ & 9.5 & P & 107 \\
\hline
 $\Delta$(1700)   & 300 & $N\pi$ & 15 & D & 580 \\
\hline
    & 300 & $\Delta\pi$ & 37.5 & S & 385 \\
\hline
$\Delta$(1910)    & 280 & $N\pi$ & 22.5 & P & 716 \\
\hline
    & 280 & $\Delta\pi$ & 60 & P & 545 \\
\hline
    & 280 & $N(1440)\pi$ & 47 & P & 393 \\
\hline
$\Delta$(1920)    & 150 & $N\pi$ & 12.5 & P & 722 \\
\hline
\end{tabular}
\caption{Nucleon and $\Delta$ particle decay data used to calculate $f_{\pi B_1 B_2}$ coupling constants. Partial widths are the average of minimum and maximum uncertainty ranges listed in PDG~\cite{Olive:2016xmw}.}
\end{center}
\label{tab:decaydata}
\end{table}

\begin{table}
\begin{center}
\begin{tabular}{|c|c|c|c|c|c|}
\hline
Parent & \begin{tabular}{@{}c@{}}Full \\ Width (MeV) \end{tabular} & \begin{tabular}{@{}c@{}}Decay \\ Mode \end{tabular} & \begin{tabular}{@{}c@{}}Partial \\ Width (\%) \end{tabular} & \begin{tabular}{@{}c@{}}Partial \\ Wave \end{tabular} & \begin{tabular}{@{}c@{}} $p_{CM}$ \\ (MeV) \end{tabular} \\
\hline
$\Lambda(1405)$ & 50 & $\Sigma \pi$ & 100 & S & 151 \\
\hline
$\Lambda(1520)$ & 16 & $\Sigma \pi$ & 42 & D & 266 \\
\hline
$\Lambda(1600)$ & 150 & $\Sigma \pi$ & 35 & P & 336 \\
\hline
$\Lambda(1670)$ & 35 & $\Sigma \pi$ & 40 & S & 393 \\
\hline
$\Lambda(1690)$ & 60 & $\Sigma \pi$ & 30 & D & 409 \\
\hline
$\Lambda(1810)$ & 150 & $\Sigma \pi$ & 25 & S & 500 \\
\hline
$\Lambda(1890)$ & 100 & $\Sigma \pi$ & 10 & P & 558 \\
\hline
$\Sigma(1385)$ & 36 & $\Lambda \pi$ & 87 & P & 208 \\
\hline
$\Sigma(1385)$ & 36 & $\Sigma \pi$ & 11.7 & P & 126 \\
\hline
$\Sigma(1670)$ & 60 & $\Sigma \pi$ & 45 & D & 415 \\
\hline
$\Xi(1530)$ & 10 & $\Xi \pi$ & 1 & P & 654 \\
\hline
\end{tabular}
\caption{Hyperon particle decay data used to calculate $f_{\pi B_1 B_2}$ coupling constants. Partial widths are the average of minimum and maximum uncertainty ranges listed in PDG~\cite{Olive:2016xmw}.}
\end{center}
\label{tab:decaydata-hyperons}
\end{table}

\begin{table}
\begin{center}
\begin{tabular}{|c|c|c|c|c|c|c|}
\hline
Parent & $A_{1/2}^p$ & $A_{1/2}^n$ & $A_{3/2}^p$ & $A_{3/2}^n$ & $p_{CM}$ (MeV) & $g_{\omega N N^{*}}$ \\
\hline
$N(1440)$ & -0.060 & 0.040 & 0 & 0 & 414 & 0.867 \\
\hline
$N(1520)$ & -0.020 & -0.050 & 0.140 & -0.115 & 470 & 2.711 \\
\hline
$N(1535)$ & 0.115 & -0.075 & 0 & 0 & 480 & 1.228 \\
\hline
$N(1650)$ & 0.045 & -0.050 & 0 & 0 & 558 & 0.193 \\
\hline
$N(1700)$ & 0.015 & 0.020 & -0.015 & -0.030 & 591 & 2.968 \\
\hline
$N(1710)$ & 0.040 & -0.040 & 0 & 0 & 597 & 0 \\
\hline
$N(1720)$ & 0.100 & -0.080 & 0.150 & -0.140 & 604 & 2.091 \\
\hline
\end{tabular}
\caption{Data used to calculate coupling constants for $\omega NN^*$ interactions derived from helicity amplitudes. Helicity amplitudes taken from PDG~\cite{Olive:2016xmw}.}
\end{center}
\end{table}

\begin{table}
\begin{center}
\subfloat{
\begin{tabular}[t]{|c|c|}
\hline
Coupling & Value \\
\hline
$f_{\pi NN}$ & 1.1  \\
\hline
$f_{\pi N \Delta}$ & 3.044 \\
\hline
$f_{N N(1440)}$ & 0.576 \\
\hline
$f_{N N(1520)}$ & 0.125 \\
\hline
$f_{N N(1535)}$ & 0.253 \\
\hline
$f_{N \Delta(1600)}$ & 0.668 \\
\hline
$f_{N \Delta(1620)}$ & 0.285 \\
\hline
$f_{N N(1650)}$ & 0.252 \\
\hline
$f_{N N(1700)}$ & 0.049 \\
\hline
$f_{N \Delta(1700)}$ & 0.109 \\
\hline
$f_{N N(1710)}$ & 0.095 \\
\hline
$f_{N N(1720)}$ & 0.243 \\
\hline
$f_{N N(1875)}$ & 0.030 \\
\hline
$f_{N N (1900)}$ & 0.138 \\
\hline
$f_{N \Delta(1910)}$ & 0.332 \\
\hline
$f_{N \Delta(1920)}$ & 0.412 \\
\hline
\end{tabular}}
\subfloat{
\begin{tabular}[t]{|c|c|}
\hline
Coupling & Value \\
\hline
$f_{\pi \Delta N(1440)}$ & 1.785  \\
\hline
$f_{\pi \Delta N(1520)}$ & 0.305  \\
\hline
$f_{\pi \Delta N(1535)}$ &0.128  \\
\hline
$f_{\pi \Delta \Delta(1600)}$ & 0.464  \\
\hline
$f_{\pi \Delta \Delta(1620)}$ & 0.158  \\
\hline
$f_{\pi \Delta N(1650)}$ & 0.135  \\
\hline
$f_{\pi \Delta N(1700)}$ & 0.368  \\
\hline
$f_{\pi \Delta \Delta(1700)}$ & 0.329  \\
\hline
$f_{\pi \Delta N(1710)}$ & 0.338  \\
\hline
$f_{\pi \Delta N(1720)}$ & 0.531  \\
\hline
$f_{\pi \Delta N(1875)}$ & 0.328  \\
\hline
$f_{\pi \Delta \Delta(1910)}$ & 0.379  \\
\hline
$f_{\pi N(1440) \Delta(1600)}$ & 4.903  \\
\hline
$f_{\pi N(1440) \Delta(1620)}$ & 0.754  \\
\hline
$f_{\pi N(1440) N(1650)}$ & 0.185  \\
\hline
$f_{\pi N(1440) \Delta(1910)}$ & 0.702  \\
\hline
$f_{\pi N(1535) N(1710)}$ & 0.487  \\
\hline
$f_{\pi N(1535) N(1900)}$ & 0.142  \\
\hline
\end{tabular}}
\subfloat{
\begin{tabular}[t]{|c|c|}
\hline
Coupling & Value \\
\hline
$f_{\Sigma \Sigma(1385)}$ & 0.683 \\
\hline
$f_{\Sigma \Lambda(1405)}$ & 1.093 \\
\hline
$f_{\Sigma \Lambda(1520)}$ & 0.216 \\
\hline
$f_{\Sigma \Lambda(1600)}$ & 0.541 \\
\hline
$f_{\Sigma \Lambda(1670)}$ & 0.222 \\
\hline
$f_{\Sigma \Sigma(1670)}$ & 0.155 \\
\hline
$f_{\Sigma \Lambda(1690)}$ & 0.131 \\
\hline
$f_{\Sigma \Lambda(1810)}$ & 0.306 \\
\hline
$f_{\Sigma \Lambda(1890)}$ & 0.251 \\
\hline
$f_{\Lambda \Sigma(1385)}$ & 0.325 \\
\hline
$f_{\Lambda \Sigma(1670)}$ & 0.110 \\
\hline
$f_{\Xi \Xi(1530)}$ & 0.654 \\
\hline\hline
$g_{\omega NN}$ & 11.0 \\
\hline
$g_{\omega N N(1440)}$ & 0.867 \\
\hline
$g_{\omega N N(1520)}$ & 2.711 \\
\hline
$g_{\omega N N(1535)}$ & 1.228 \\
\hline
$g_{\omega N N(1650)}$ & 0.445 \\
\hline
$g_{\omega N N(1700)}$ & 0.193 \\
\hline
$g_{\omega N N(1710)}$ & 0.0 \\
\hline
$g_{\omega N N(1720)}$ & 2.091 \\
\hline
$g_{\omega N N(1875)}$ & 2.015 \\
\hline
$g_{\omega N N(1900)}$ & 2.887 \\
\hline
\end{tabular}}
\caption{Baryonic coupling constants used in this work.}
\label{tab:couplings}
\end{center}
\end{table}

\begin{flushleft}

\end{flushleft}

\begin{thebibliography}{10}

\bibitem{Rapp:1999ej}
  R.~Rapp and J.~Wambach,
  Adv.\ Nucl.\ Phys.\  {\bf 25}, 1 (2000).

\bibitem{Arnaldi:2008er}
R.~Arnaldi \textit{et al.} [NA60],
Eur. Phys. J. C \textbf{59} (2009), 607-623.

\bibitem{vanHees:2011vb}
  H.~van Hees, C.~Gale and R.~Rapp,
  Phys.\ Rev.\ C {\bf 84}, 054906 (2011).

\bibitem{Adare:2014fwh}
  A.~Adare {\it et al.} [PHENIX Collaboration],
  Phys.\ Rev.\ C {\bf 91}, no. 6, 064904 (2015).

\bibitem{Adam:2015lda}
  J.~Adam {\it et al.} [ALICE Collaboration],
  Phys.\ Lett.\ B {\bf 754}, 235 (2016).

\bibitem{Adare:2015lcd}
  A.~Adare {\it et al.} [PHENIX Collaboration],
  Phys.\ Rev.\ C {\bf 94}, no. 6, 064901 (2016).

\bibitem{Acharya:2018bdy}
  S.~Acharya {\it et al.} [ALICE Collaboration],
  Phys.\ Lett.\ B {\bf 789}, 308 (2019).

\bibitem{vanHees:2014ida}
  H.~van Hees, M.~He and R.~Rapp,
  Nucl.\ Phys.\ A {\bf 933}, 256 (2015).

\bibitem{Paquet:2015lta}
  J.~F.~Paquet, C.~Shen, G.~S.~Denicol, M.~Luzum, B.~Schenke, S.~Jeon and C.~Gale,
  Phys.\ Rev.\ C {\bf 93}, no. 4, 044906 (2016).



\bibitem{Rarita:1941mf}
  W.~Rarita and J.~Schwinger,
  Phys.\ Rev.\  {\bf 60}, 61 (1941).

\bibitem{Weinberg:1996kr}
  S.~Weinberg,
  ``\textit{The Quantum Theory of Fields. Vol. 2: Modern Applications,}''
  Cambridge University Press (New York, 2005).

\bibitem{Gasparyan:2003fp}
  A.~M.~Gasparyan, J.~Haidenbauer, C.~Hanhart and J.~Speth,
  Phys.\ Rev.\ C {\bf 68}, 045207 (2003).

\bibitem{Wess:1971yu}
  J.~Wess and B.~Zumino,
  Phys.\ Lett.\  {\bf 37B}, 95 (1971).

\bibitem{Witten:1983tx}
  E.~Witten,
  Nucl.\ Phys.\ B {\bf 223}, 433 (1983).

\bibitem{Sakurai}
  J.J.~Sakurai,
  ``\textit{Currents and Mesons,}''
  The University of Chicago Press (Chicago, 1969).

\bibitem{Ericson:1988gk}
  T.~E.~O.~Ericson and W.~Weise,
  Int.\ Ser.\ Monogr.\ Phys.\  {\bf 74} (1988).

\bibitem{Machleidt:1989tm}
  R.~Machleidt,
  Adv.\ Nucl.\ Phys.\  {\bf 19}, 189 (1989).

\bibitem{Bieniek:2001xu}
  A.~Bieniek, A.~Baran and W.~Broniowski,
  Phys.\ Lett.\ B {\bf 526}, 329 (2002).

\bibitem{Turbide:2003si}
  S.~Turbide, R.~Rapp and C.~Gale,
  Phys.\ Rev.\ C {\bf 69}, 014903 (2004).

\bibitem{Holt:2015cda}
  N.~P.~M.~Holt, P.~M.~Hohler and R.~Rapp,
  Nucl.\ Phys.\ A {\bf 945}, 1 (2016).

\bibitem{Velo:1970ur}
  G.~Velo and D.~Zwanziger,
  Phys.\ Rev.\  {\bf 188}, 2218 (1969).

\bibitem{Benmerrouche:1989uc}
  M.~Benmerrouche, R.~M.~Davidson and N.~C.~Mukhopadhyay,
  Phys.\ Rev.\ C {\bf 39}, 2339 (1989).

\bibitem{Urban:1998eg}
  M.~Urban, M.~Buballa, R.~Rapp and J.~Wambach,
  Nucl.\ Phys.\ A {\bf 641}, 433 (1998).

\bibitem{Urban:1999im}
  M.~Urban, M.~Buballa, R.~Rapp and J.~Wambach,
  Nucl.\ Phys.\ A {\bf 673}, 357 (2000).

\bibitem{Rapp:1999us}
  R.~Rapp and J.~Wambach,
  Eur.\ Phys.\ J.\ A {\bf 6}, 415 (1999).

\bibitem{Riek:2008ct}
  F.~Riek, R.~Rapp, T.-S.~H.~Lee and Y.~Oh,
  Phys.\ Lett.\ B {\bf 677}, 116 (2009).

\bibitem{Herrmann:1993za}
  M.~Herrmann, B.~L.~Friman and W.~Norenberg,
  Nucl.\ Phys.\ A {\bf 560}, 411 (1993).

\bibitem{Mathiot:1984bs}
  J.~F.~Mathiot,
  Nucl.\ Phys.\ A {\bf 412}, 201 (1984).

\bibitem{Roth}
  T.~Roth, Dimplomarbeit, Technischen Universit{\"a}t Darmstadt (1999).


\bibitem{Rapp:1999qu}
  R.~Rapp and C.~Gale,
  Phys.\ Rev.\ C {\bf 60}, 024903 (1999).

\bibitem{Olive:2016xmw}
  C.~Patrignani {\it et al.} [Particle Data Group],
  Chin.\ Phys.\ C {\bf 40}, no. 10, 100001 (2016).

\bibitem{Halzen:1984mc}
  F.~Halzen and A.~D.~Martin,
  ``\textit{Quarks And Leptons: An Introductory Course In Modern Particle Physics,}''
  John Wiley \& Sons (New York, 1984).

\bibitem{Rapp:1997fs}
  R.~Rapp, G.~Chanfray and J.~Wambach,
  Nucl.\ Phys.\ A {\bf 617}, 472 (1997).

\bibitem{Post:2000rf}
  M.~Post and U.~Mosel,
  Nucl.\ Phys.\ A {\bf 688}, 808 (2001).

\bibitem{Armstrong:1971ns}
  T.~A.~Armstrong {\it et al.},
  Phys.\ Rev.\ D {\bf 5}, 1640 (1972).

\bibitem{Bartalini:2008zza}
  O.~Bartalini {\it et al.},
  Phys.\ Atom.\ Nucl.\  {\bf 71}, 75 (2008).

\bibitem{Lenz:1975qx}
  F.~Lenz and E.~J.~Moniz,
  Phys.\ Rev.\ C {\bf 12}, 909 (1975).

\bibitem{Woloshyn:1976ca}
  R.~M.~Woloshyn, E.~J.~Moniz and R.~Aaron,
  Phys.\ Rev.\ C {\bf 13}, 286 (1976).

\bibitem{Moniz:1981zz}
  E.~J.~Moniz and A.~Sevgen,
  Phys.\ Rev.\ C {\bf 24}, 224 (1981).

\bibitem{Brown:1975di}
  G.~E.~Brown and W.~Weise,
  Phys.\ Rept.\  {\bf 22}, 279 (1975).

\bibitem{Chung:1995dx}
  S.~U.~Chung, J.~Brose, R.~Hackmann, E.~Klempt, S.~Spanier and C.~Strassburger,
  Annalen Phys.\  {\bf 4}, 404 (1995).

\bibitem{Oset:1981ih}
  E.~Oset, H.~Toki and W.~Weise,
  Phys.\ Rept.\  {\bf 83}, 281 (1982).

\bibitem{Workman:2012hx}
  R.~L.~Workman, R.~A.~Arndt, W.~J.~Briscoe, M.~W.~Paris and I.~I.~Strakovsky,
  Phys.\ Rev.\ C {\bf 86}, 035202 (2012).


\bibitem{Rapp:2000pe}
R.~Rapp,
Phys. Rev. C \textbf{63}, 054907 (2001).



\end{thebibliography}
\end{document}